\documentclass[journal,10pt,twocolumn]{IEEEtran}
\normalsize
\usepackage{amsmath, amsthm, amssymb}
\usepackage{epsfig}
\usepackage{latexsym}
\usepackage{epstopdf}
\usepackage{multirow}
\usepackage{graphicx}
\usepackage{cite}
\usepackage{caption}
\usepackage[linesnumbered,ruled]{algorithm2e}
\usepackage{algpseudocode}
\usepackage{siunitx}
\usepackage{float}

\newcommand\floor[1]{\lfloor#1\rfloor}

\algnewcommand\algorithmicinput{\textbf{Input:}}
\algnewcommand\INPUT{\item[\algorithmicinput]}
\algnewcommand\algorithmicoutput{\textbf{Output:}}
\algnewcommand\OUTPUT{\item[\algorithmicoutput]}
\begin{document}
\title{Joint Uplink/Downlink Optimization for Backhaul-Limited Mobile Cloud Computing with  User Scheduling}

\author{Ali Al-Shuwaili, Osvaldo Simeone, Alireza Bagheri, and Gesualdo Scutari }

\maketitle


\begin{abstract}
Mobile cloud computing enables the offloading of computationally heavy applications, such as for gaming, object recognition or video processing, from mobile users (MUs) to cloudlet or cloud servers, which are connected to wireless access points, either directly or through finite-capacity backhaul links. 
In this paper, the design of a mobile cloud computing system is investigated by proposing the joint optimization of computing and communication
resources with the aim of minimizing the energy required for offloading across all MUs under latency constraints at the application layer. The proposed design  accounts for multi-antenna uplink and downlink interfering transmissions, with or without cooperation on the downlink, along with the allocation of backhaul and computational resources and user selection. The resulting design optimization problems are non-convex, and stationary solutions are computed  by means of successive convex approximation (SCA) techniques. Numerical results illustrate the advantages in terms of energy-latency trade-off of the joint optimization of computing and communication resources, as well as the impact of system parameters, such as backhaul capacity, and of the network architecture.
\end{abstract}


\let\thefootnote\relax\footnotetext{A. Al-Shuwaili, O. Simeone, and A. Bagheri are with the Center for Wireless Information Processing (CWiP),  Department of Electrical and Computer Engineering, New Jersey Institute of Technology, Newark, NJ 07102 USA (e-mail: ana24@njit.edu, osvaldo.simeone@njit.edu, and ab745@njit.edu).

 G. Scutari is with the Department of Industrial Engineering, Purdue
University, West Lafayette, IN 47907 USA (e-mail: gscutari@purdue.edu).

This work was partially supported by the U.S. NSF through grant 1525629.
}

\section{Introduction}\label{sec:intro}

Mobile cloud computing enables the offloading of computationally heavy applications from Mobile Users (MUs) to a remote server by means of a wireless communication network \cite{kumar2013survey, dinh2013survey,sanaei2014heterogeneity, lei2013challenges, cloudlet}. Given the battery-limited nature of mobile devices, mobile cloud computing makes it possible to provide services, such as gaming, object recognition, video processing, or augmented reality, that may otherwise not be available to mobile users. Offloading may take place to remote servers in the ``cloud'', which is accessed by the wireless access points via backhaul links, or to ``cloudlet'' servers that are directly connected to an access point \cite{cloudlet}. In the latter case, the approach is also known as mobile edge computing, which is currently seen as a key enabler of the so called tactile internet \cite{tac} and is subject to standardization efforts \cite{mec}. Offloading to peer mobile devices is also being considered \cite{m2m}. Examples of architectures that are based on mobile cloud/cloudlet computation offloading include MAUI \cite{maui}, ThinkAir \cite{kosta2012thinkair}, and MobiCoRE \cite{mobcore}, while examples of commercial application based on mobile cloud computing include Apple iCloud, Shazam, and Google Goggles \cite{MCCAPP, shazam}.

When optimized solely at the application layer, as in most of the earlier literature, the design of a mobile cloud computing systems entail decision of whether a mobile can decrease its energy expenditure by offloading the execution of the entire application \cite{kumar2013survey,dinh2013survey,huang2012dynamic,Miettinen2010EE,kumar2010cloud} or of some selected subtasks (see, e.g., \cite{Hermes,Barbarossa1}). Given that offloading requires transmission and reception on the wireless interface, a more systematic approach involves the joint optimization of offloading decisions and communication parameters, such as uplink power allocation \cite{Barbarossa1,shahrooz}, link and subcarrier selection \cite{xiang2014energy,2016letaief}, and uplink data rate \cite{vidal, GE}. 

The joint optimization of the set of subtasks to be offloaded and of the uplink transmission powers was studied in \cite{Barbarossa1,shahrooz} under static channel condition. A related dynamic computation offloading approach was presented in \cite{huang2012dynamic, GE} that  determines  which  application subtasks are to be executed remotely given the available wireless network connectivity. Assuming a simplified execution model, in which input data can be arbitrary partitioned for separate processing, references  \cite{vidal, DVS} study  the fraction of the mobile data to be offloaded to the cloud, with the rest being executed locally at the mobile device. A framework that integrates wireless energy transfer and cloud mobile computing was put forth in \cite{MPT}.  

The works summarized thus far focus on the operation of a \textit{single} MU. In contrast to the single-MU problem formulation, in a scenario with multiple MUs transmitting over a shared wireless medium across multiple cells, the design of a mobile cloud computing system requires: (\emph{i}) the management of interference for the \textit{uplink}, through which MUs offload the data needed for computation in the cloud; (\emph{ii}) the management of interference for the \textit{downlink}, through which the outcome of the cloud computations are fed back to the MUs; (\emph{iii}) the allocation of \textit{backhaul} resources for communication between wireless edge and cloud; and (\emph{iv}) the allocation of \textit{computing} resources at the cloud. Furthermore, the optimization should include \textit{user selection}, or \textit{scheduling} mechanisms whereby the offloading users are guaranteed an energy consumption that is smaller than the amount required for local computing at the device.

The limited literature on resource
allocation and offloading decisions for the multiuser case includes papers \cite{xchen2015, chen2015,hossain2013, sar02, you2016,molina2014,zhang2016, guo2016,  zhao2015}. The problem of decentralized user scheduling when modeling the aspect (\emph{i}) of uplink interference is considered in \cite{xchen2015,chen2015} for single-antenna elements within a game-theoretic framework. The joint allocation of radio and computational resources is considered in \cite{hossain2013} by accounting for the elements (\emph{i}) and (\emph{iv}), in the presence of multiple clouds,  with the aim of maximizing network operator revenue via resource pool sharing.  A problem formulation including elements (\emph{i}) and (\emph{iv}) was studied in \cite{sar02}  with MIMO transceivers and for a fixed set of scheduled users. Scheduling in a single-cell was considered in \cite{you2016} with the goal of  minimizing the weighted sum mobile energy  consumption; it was shown that the  optimal  scheduling and cloud resource allocation  policy (element (\emph{iv})) have a threshold-based structure. Another scheduling strategy for multiuser offloading systems in a small-cell set-up is presented in \cite{molina2014},   where the resources are allocated under the objective of minimizing the average  latency experienced by the worst-case user by accounting for element (\emph{iv}) with the inclusion of uplink and downlink tasks schedulers. A scheme that jointly optimizes the computation offloading decisions and the radio resource allocation in heterogeneous networks by accounting for element (\emph{i})  so as to minimize the mobile energy expenditure under latency constraints was proposed in \cite{zhang2016}. An energy-efficient resource allocation for interference-free multi-users  scenario was discussed in \cite{guo2016} with the aim of optimizing uplink and downlink transmissions duration while considering element (\emph{iv}). Finally, the problem of scheduling tasks between cloud and edge processors was studied in  \cite{zhao2015} without modeling the physical layer.

In this paper, we account for all elements (\emph{i})-(\emph{iv}) as well as for user scheduling. Specifically, our main contributions are:

\begin{itemize}
\item We study the problem of minimizing the mobile energy consumption under latency constraints over uplink and downlink precoding, uplink and downlink backhaul resource allocation, as well as cloud computing resource allocation for general multi-antenna transceivers. This work is mostly motivated by the increasing importance of backhaul capacity limitations, which are well understood to be often the bottleneck in modern dense network deployments (see, e.g.,\cite{g5,bh}). We emphasize that, unlike  \cite{sar02}, the problem formulation explicitly models the optimization of downlink communication for downloading the outcome of the optimization at the MUs as well as of the backhaul resource allocated to the active MUs in both uplink and downlink, that is, it accounts for all elements (\emph{i})-(\emph{iv}) listed above. The resulting problem requires the development of a novel adaptation of the Successive Convex Approximation (SCA) scheme \cite{scutari2014distributed, scutari2016parallel}  that accounts for downlink and backhaul transmissions.
\item
The optimization of users scheduling is tackled jointly with the operation of uplink/downlink, backhaul and computational resources under the key constraint that each offloading MU should not consume more energy than that required for local computation. The resulting mixed integer problem is tackled by a means of SCA coupled with the smooth $l_p$-norm approximation approach \cite {ran}. We emphasize that this problem was not studied in [30], which instead considered a fixed set of scheduled users.

\item
A hybrid cloud-edge computing setup is studied in which, beside a cloud server, ``cloudlet'' or ``edge'' servers are available locally at the wireless access points.  The cloudlet servers are able to execute offloaded applications without incurring backhaul latency but with a generally smaller CPU frequency \cite{cloudlet}. For the first time, we investigate here the optimal task allocation between cloudlet and the cloud via SCA in the presence of all elements (\emph{i})-(\emph{iv}) discussed above.  
 
 \item 
 We study the impact of cooperative downlink transmission via network MIMO \cite {comp} on the achievable  energy-latency trade-off by accounting for the backhaul overhead needed to deliver user data to multiple access points for transmission to the MUs. This has also not previously studied in the context of mobile cloud computing with backhaul limitations.
\end{itemize}

The rest of the paper is organized as follows. Section II introduces the system model along with the basic problem formulation. The proposed SCA solution is described in Section III.  Users scheduling is studied in Section IV. Hybrid edge, or cloudlet, and cloud computing is formulated and tackled in Section V. Cooperative downlink transmission is discussed in Section VI. Numerical results are presented in Section VII, while concluding remarks are finally provided in Section VIII.

\section {System Model and Problem Formulation} \label{sec:problem}
In this section, we describe the basic system model and problem formulation that will be adopted in this paper. The basic model assumes fixed user scheduling, offloading to a centralized cloud processor, and non-cooperative transmission at the wireless access points. Generalizations that address user scheduling, local computing capabilities at the access points and cooperative transmission will be treated later in Section \ref{selection}, Section \ref{sec:hybrid} and Section \ref{netmimo}, respectively.
\subsection {System Model}
We consider a network composed of $N_c$ cells of possibly different sizes such as micro- or femto-cells. 
%
\begin{figure}[!t] \label{fig1}
        \centering
        \includegraphics[width=\columnwidth]{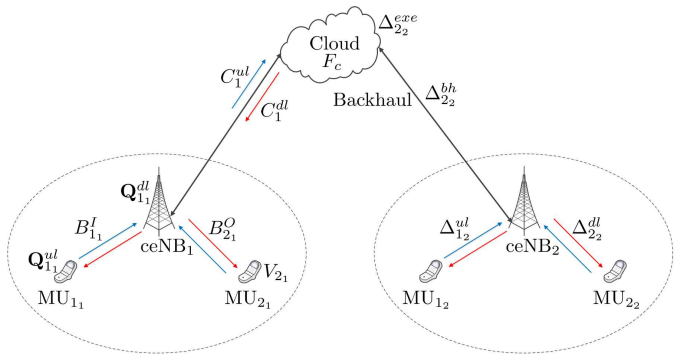}
        \caption{{\small{Basic system model: Mobile users (MUs) offload the execution of their applications to a centralized cloud processor through a wireless access network and finite capacity backhaul links.}}}
        \label{sys}
\end{figure}
%
Each cell $n=1,\ldots,N_c$ includes a  base station, referred to as cloud-enhanced e-Node B (ceNB) borrowing from LTE nomenclature, which is connected to a common cloud server that provides computational resources. As shown in Fig. 1, each cell contains $K$ active mobile users (MUs) that have been scheduled for the offloading of their local applications to the cloud processor. The $K$ MUs in the same cell transmit in orthogonal spectral resources, in  the time or frequency domain. We denote by $i_n$ the MU in cell $n$ that is scheduled on the $i$-th spectral resource, and by $\mathcal{I} \triangleq \left\{i_n: i=1,\ldots,K,    n=1,\ldots,N_c \right\}$ the set of all the active MUs in the system. Each MU $i_n$ and ceNB $n$ is equipped with $N_{T_{i_n}}$ transmit and $N_{R_{n}}$ receive antenna, respectively. Note that MUs in   different cells that are scheduled on the same spectral resources interfere with each other.

Each MU $i_n$ wishes to run an application within a given maximum latency $T_{{i}_n}$. The application to be executed is characterized by the number $V_{{i}_n}$ of CPU cycles necessary to complete it, by the number $B_{{i}_n}^I$ of input bits, and by the number $B^O_{{i}_n}$ of output bits encoding the result of the computation. Table I summarizes the notations and parameters used in the system model.

 We next derive energy and latency resulting from offloading of the applications of all active MUs. The offloading latency consist of the time $\Delta_{i_{n}}^{ul}$ needed for the MU to transmit the input bits to its ceNB in the uplink; the time $\Delta_{i_{n}}^{exe}$  necessary for the cloud to execute the instructions; the round-trip time $\Delta_{i_{n}}^{bh}$ for exchanging information between ceNB and the cloud through the backhaul link; and the time $\Delta_{i_{n}}^{dl}$ to send the result back to the MU in the downlink (see Fig. 2). We can hence write the total offloading latency for MU $i_n$ as
 \begin{equation}
 \Delta_{i_{n}}=\Delta_{i_{n}}^{ul}+\Delta_{i_{n}}^{exe}+\Delta_{i_{n}}^{bh}+\Delta_{i_{n}}^{dl}.
 \end{equation}
The energy $E_{i_{n}}$ of each MU $i_n$ instead depends only on the power used for transmission in the uplink and reception energy in the downlink. These latency and energy terms are computed as a function of the radio and computational resources as detailed next.
\begin{figure}[!t] \label{fig2}
        \centering
        \includegraphics[width=\columnwidth]{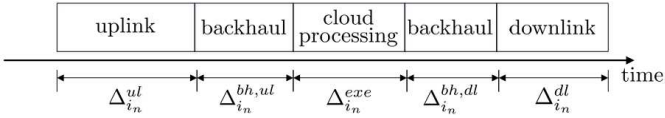}
        \caption{{\small{Timeline of offloading from an MU $i_n$. Note that the total two-way backhaul latency is given as $\Delta_{i_{n}}^{bh} = \Delta_{i_{n}}^{bh,ul} + \Delta_{i_{n}}^{bh,dl}$. }}}
        \label{timeline}
\end{figure}

1) \textit{Uplink transmission:} The optimization variables at the physical layer for the uplink  are the users' transmit covariance matrices $\mathbf{Q}^{ul} \triangleq \left({\mathbf{Q}}_{i_{n}}^{ul}\right)_{i_{n}\in\mathcal{I}},$ where ${\mathbf{Q}}_{i_{n}}^{ul}=\mathbb{E}[{\mathbf{x}}_{i_{n}}^{ul}{\mathbf{x}}_{i_{n}}^{{ul}^H}]$ with $\mathbf{x}_{i_{n}}^{ul}\sim \mathcal{CN}(\mathbf{0},{\mathbf{Q}}_{i_{n}}^{ul})$ being the signal transmitted by the user $i_n$. These matrices are subject to power budget constraints
\begin{equation}
\mathcal{Q}_{i_{n}}^{ul} \triangleq \left\{{\mathbf{Q}}_{i_{n}}^{ul} \in {\mathbb{C}}^{N_{{T}_{i_{n}}}\times N_{{T}_{i_{n}}} } :{\mathbf{Q}}_{i_{n}}^{ul}\succeq 0, \text{tr}({\mathbf{Q}}^{ul}_{i_{n}}) \leq P_{i_{n}}^{ul} \right\},
\end{equation}
where $P_{i_{n}}^{ul}$ is the maximum allowed transmit energy per symbol of MU $i_n$. For any given profile $\mathbf{Q}^{ul}$, the achievable transmission rate, in bits per symbol, which corresponds to the mutual information evaluated on the MIMO Gaussian channel (see, e.g., \cite{tse2005fundamentals}) of MU $i_n$ when  multi-user interference is treated as additive Gaussian noise,   is
\begin{equation}
r_{i_{n}}^{ul}(\mathbf{Q})=\log_{2}{\det}\Big(\mathbf{I}+\mathbf{H}_{i_{n}}^H\mathbf{R}_n^{ul}(\mathbf{Q}_{-i_{n}}^{ul})^{-1}\mathbf{H}_{i_{n}}{\mathbf{Q}}_{i_{n}}^{ul}\Big),
\end{equation}
where
\begin{equation}
\mathbf{R}_n^{ul}(\mathbf{Q}_{-i_{n}}^{ul})\triangleq N_0\mathbf{I}+ \sum_{j_m\in \mathcal{I},m\neq n}\mathbf{H}_{j_{m}n}\mathbf{Q}_{j_{m}}^{ul}\mathbf{H}_{j_{m}n}^H
\end{equation}
is the covariance matrix of the sum of the noise and of the inter-cell interference affecting reception at the $n$-th ceNB in the $i$-th spectral resources; $\mathbf{Q}_{-i_{n}}^{ul} \triangleq \left( \left(\mathbf{Q}_{j_{m}}^{ul}\right)_{j=1}^K \right)_{n \ne m=1}^{N_c}$; $N_0$ is the noise power spectral density; and $\mathbf{H}_{i_{n}}$ is the uplink channel matrix for MU $i_n$ to the ceNB in the cell $n$, whereas $\mathbf{H}_{j_{m}n}$ is the cross channel matrix between the interfering MU $j_m$ in the cell $m$ and the ceNB in cell $n$. The channel matrices account for path loss, slow and fast fading. The time, in seconds, necessary for user $i$ in cell $n$ to transmit the input bits $B_{i_{n}}^I$
 to its ceNB in the uplink is then
\begin{equation} \label{del_ul}
\Delta _{{i_n}}^{ul}\left( {{{\bf{Q}}^{ul}}} \right) = \frac{{B_{{i_n}}^I}}{{{W^{ul}}r_{{i_n}}^{ul}\left( {{{\bf{Q}}^{ul}}} \right)}},
\end{equation}
where $W^{ul}$ is the uplink channel bandwidth allocated to each one of the orthogonal spectral resources. The corresponding energy consumption due to uplink transmission is then
\begin{equation} \label{off_ene}
{E_{{i_n}}^{ul}}\left( {{{\bf{Q}}^{ul}}} \right) = {B_{{i_n}}^I}\frac{{\text{tr}\left( {{\bf{Q}}_{{i_n}}^{ul}} \right)}}{{r_{{i_n}}^{ul}\left( {{{\bf{Q}}^{ul}}} \right)}},
\end{equation}
since around ${{B_{{i_n}}^I}}/{{{r_{{i_n}}^{ul}\left( {{{\bf{Q}}^{ul}}} \right)}}}$ channel uses are needed  in order to transmit reliably in the uplink.

\begin{table}[]
\centering
\caption{System Parameters}
\label{my-label}
\begin{tabular}{|l|l|}
\hline
Parameter    & Description \\ \hline
$F_c$, $F^{ceNB}_n$      & cloud and cloudlet computational capacity    \\ \hline
 $C^{ul}_{n},C^{dl}_{n}$         & uplink and downlink backhaul capacity of cell $n$        \\ \hline
$H_{{i}_n},G_{{i}_n}$       & uplink and downlink channel matrix for user $i_n$     \\ \hline
$P_{i_{n}}^{ul},P_{n}^{dl}$ & uplink and downlink power budget constraint      \\ \hline
$W^{ul},W^{dl}$       & uplink and downlink bandwidth           \\ \hline
$N_o$ & noise power spectral density      \\ \hline
$N_{T_{i_n}},N_{R_{n}}$ & number of transmit and receive antenna \\ \hline
$B_{{i}_n}^I,B^O_{{i}_n} $ & number of input and output bits for user $i_n$ \\ \hline
$V_{{i}_n}$ & number of CPU cycles for user $i_n$\\ \hline
$T_{{i}_n}$ & latency constraint for user $i_n$ \\ \hline
$N_c, K$ & number of cell and number of users in each cell \\ \hline
$d_{{i}_n}$ & reception energy constant for user $i_n$ \\ \hline
$\kappa$ & mobile device switched capacitance \\ \hline
$\alpha, \delta$ & step size constant and termination accuracy \\ \hline

\end{tabular}
\end{table}
2) \textit{Downlink transmission:} The optimization variables for the downlink are the ceNBs' transmit covariance matrices $\left(\mathbf{Q}_{i_{n}}^{dl}\right)_{i=1}^K$, which are subject to per-ceNB power constraints $P^{dl}_n$:
\begin{equation}
\resizebox{ \columnwidth}{!} {
$\mathcal{Q}_{{n}}^{dl} \triangleq \left\{\left(\mathbf{Q}_{i_{n}}^{dl}\right)_{i=1}^K \in {\mathbb{C}}^{N_{{T}_{i_{n}}}\times N_{{T}_{i_{n}}} } : {\mathbf{Q}}_{i_{n}}^{dl}\succeq 0,  \displaystyle\sum_{i=1}^K\text{tr}({\mathbf{Q}}_{i_{n}}^{dl}) \leq P^{dl}_n \right\}.$ }
\end{equation}
Similar to the uplink, we can write the achievable rate in bits per symbol for each MU in the downlink as
\begin{equation} \label{r_dl}
r_{i_{n}}^
{dl}(\mathbf{Q}^
{dl})=\log_{2}{\det}\Big(\mathbf{I}+ \mathbf{G}_{i_{n}}^H  \mathbf{R}_n^{dl}(\mathbf{Q}_{-i_{n}}^
{dl})^{-1}\mathbf{G}_{i_{n}}{\mathbf{Q}}_{i_{n}}^{dl}\Big),
\end{equation}
with
\begin{equation} \label{noise_dl}
\mathbf{R}_n^{dl}(\mathbf{Q}_{-i_{n}}^
{dl})\triangleq N_0\mathbf{I}+ \sum_{j_m\in \mathcal{I},m\neq n}\mathbf{G}_{j_{m}n}\mathbf{Q}_{j_{m}}^{dl}\mathbf{G}_{j_{m}n}^H;
\end{equation}
  the corresponding required transmission time reads
\begin{equation} \label{del_dl}
\Delta _{{i_n}}^{dl}\left( {{{\bf{Q}}^{dl}}} \right) = \frac{{B_{{i_n}}^O}}{{{W^{dl}}r_{{i_n}}^{dl}\left( {{{\bf{Q}}^{dl}}} \right)}},
\end{equation}
where $\mathbf{G}_{i_{n}}$ is the downlink  channel matrix between the ceNB in cell $n$ and the MU $i_n$; $\mathbf{G}_{j_{m}n}$
is the cross channel matrix between the interfering MU $j_m$ in the cell $m$ and the ceNB in cell $n$; ${W^{dl}}$ is the downlink channel bandwidth; and $\mathbf{Q}_{-i_{n}}^{dl} \triangleq \left( \left(\mathbf{Q}_{j_{m}}^{dl}\right)_{j=1}^K \right)_{n \ne m=1}^{N_c}$. The downlink energy consumption is given by
\begin{equation} \label{dl_ene}
{E_{{i_n}}^{dl}}\left( {{{\bf{Q}}^{dl}}} \right) = {B_{{i_n}}^O}\frac{{d_{{i_n}}}}{{r_{{i_n}}^{dl}\left( {{{\bf{Q}}^{dl}}} \right)}},
\end{equation}
where ${d_{{i_n}}}$ is parameter that indicates the receiver energy expenditure for each symbol interval.
Note that (7)-(8) implicitly assume that the downlink spectral resources are allocated to the MUs in the same way as for the uplink, so that MUs ${i_{n}} \text{ for } n=1,\ldots,N_c$ are mutually interfering in both uplink and downlink. This assumption can be easily alleviated at the cost of introducing additional notation.
\newtheorem{remark}{Remark}

3) \textit{Cloud processing:} Let $F_c$ be the capacity in terms of number of CPU cycles per second of the cloud; and let $f_{i_{n}}\geq 0$ be the fraction of the processing power $F_c$ assigned to user $i_n$, so that $\sum_{{i_n}\in\mathcal{I}} f_{i_{n}}\leq 1$.  The time needed to run $V_{i_{n}}$ CPU cycles for user $i_n$ remotely is then
\begin{equation} \label{del_exe}
\Delta_{i_{n}}^{exe}(f_{i_{n}})=\frac{V_{i_{n}}}{f_{i_{n}}F_c}.
\end{equation}
Define $\mathbf{f} \triangleq ({{f}}_{i_{n}})_{i_{n}\in\mathcal{I}}$.

4) \textit{Backhaul transmission:} We denote by  $C_{{n}}^{ul}$ the capacity in bits per second of the backhaul connecting the ceNB in cell $n$ with the cloud, and by $C_{{n}}^{dl}$  the capacity in bits per second of the backhaul connecting the cloud with the ceNB in cell $n$. Let $  c_{i_{n}}^{ul},c_{i_{n}}^{dl} \geq 0$ be the fraction of the backhaul capacities $C_{{n}}^{ul}$ and $C_{{n}}^{dl}$, respectively, allocated to MU $i$ in cell $n$. We then have the constraint $\sum_{i=1}^K c_{i_{n}}^{ul}\leq 1$ and $\sum_{i=1}^K c_{i_{n}}^{dl}\leq 1$ for all $n$. Moreover, the time delay due to the backhaul transfer between ceNB $n$ and the cloud in both directions is given by
\begin{equation} \label{del_bh}
\Delta_{i_{n}}^{bh}(c_{i_{n}}^{ul},c_{i_{n}}^{dl})=\frac{B_{i_{n}}^I}{c_{i_{n}}^{ul} C_{n}^{ul}}+\frac{B_{i_{n}}^O}{c_{i_{n}}^{dl}C_{n}^{dl}},
\end{equation}
where the first term represents the latency in the uplink direction, denoted as $\Delta_{i_{n}}^{bh,ul}$ (cf. Fig. 2), and the second term represents the latency in the downlink direction, denoted as $\Delta_{i_{n}}^{bh,dl}$ in the same figure. The uplink/downlink backhaul allocation vectors are defined as $\mathbf{c}^{ul} \triangleq ({{c}}_{i_{n}}^{ul})_{i_{n}\in\mathcal{I}}$ and $\mathbf{c}^{dl} \triangleq ({{c}}^{dl}_{i_{n}})_{i_{n}\in\mathcal{I}}$, respectively.
\subsection{Problem Formulation}
The total energy consumption due to the offloading of user $i_n$ data, i.e., transmitting $B_{i_{n}}^I$ bits and receiving $B_{i_{n}}^O$ bits, is then given by
\begin{equation} \label{mu-ene}
\begin{split}
& E_{{i_n}} \left( {{{\bf{Q}}^{ul}}}, {{{\bf{Q}}^{dl}}}\right) = {E_{{i_n}}^{ul}}\left( {{{\bf{Q}}^{ul}}} \right) + {E_{{i_n}}^{dl}}\left( {{{\bf{Q}}^{dl}}} \right)  \\
 & =  {B_{{i_n}}^I}\frac{{\text{tr}\left( {{\bf{Q}}_{{i_n}}^{ul}} \right)}}{{r_{{i_n}}^{ul}\left( {{{\bf{Q}}^{ul}}} \right)}}+{B_{{i_n}}^O}\frac{{d_{{i_n}}}}{{r_{{i_n}}^{dl}\left( {{{\bf{Q}}^{dl}}} \right)}}.
\end{split}
\end{equation}
The optimal offloading problem can be stated as the minimization of the sum of the energy  spent by all MUs to run their applications remotely, subject to individual latency and power constraint. Stated in mathematical terms, we have the following

\begin{equation*}
\begin{array}{l}
\begin{array}{*{20}{l}}
{\mathop {{\mathop{\rm min}\nolimits} }\limits_{{{\bf{Q}}^{ul}},{{\bf{Q}}^{dl}},{\bf{f}},{{\bf{c}}^{ul}},{{\bf{c}}^{dl}}} }&{E\left( {{{\bf{Q}}^{ul}}},{{{\bf{Q}}^{dl}}} \right)}{ \triangleq \sum\limits_{{i_n} \in {\cal I}} {{E_{{i_n}}}\left( {{\bf{Q}}^{ul},{\bf{Q}}^{dl}} \right)} }\\
{}{}&{ = \sum\limits_{{i_n} \in {\cal I}} {B_{{i_n}}^I}\frac{{\text{tr}\left( {{\bf{Q}}_{{i_n}}^{ul}} \right)}}{{r_{{i_n}}^{ul}\left( {{{\bf{Q}}^{ul}}} \right)}}+{B_{{i_n}}^O}\frac{{d_{{i_n}}}}{{r_{{i_n}}^{dl}\left( {{{\bf{Q}}^{dl}}} \right)}} }
\end{array}\\
\begin{array}{*{20}{l}}
\;\;\;\;\;\;\;\;\; \text{s.t.}&{{\bf{C}}.\mathbf{1}}&{\frac{{B_{{i_n}}^I}}{{{W^{ul}}r_{{i_n}}^{ul}\left( {{{\bf{Q}}^{ul}}} \right)}} + \frac{{B_{{i_n}}^I}}{{c_{{i_n}}^{ul}{C^{ul}_n}}} + \frac{{{V_{{i_n}}}}}{{{f_{{i_n}}}{F_c}}}}\\
{}&{}&{ + \frac{{B_{{i_n}}^O}}{{c_{{i_n}}^{dl}{C^{dl}_n}}} + \frac{{B_{{i_n}}^O}}{{{W^{dl}}r_{{i_n}}^{dl}\left( {{{\bf{Q}}^{dl}}} \right)}} \le {T_{{i_n}}}}, \forall i_n \in \cal{I},\\
{}&{{\bf{C}}{\bf{.2}}}&{{f_{{i_n}}} \ge 0}, \forall i_n \in {\cal{I}}, \sum\limits_{{i_n} \in {\mathcal {I}}} {{f_{{i_n}}}}  \leq 1, \;\;\;\;\;\;\;\;\;\;\;\;\;\;\;\;\;\;\;\; \\
{}&{{\bf{C}}{\bf{.3}}} &  { c_{{i_n}}^{ul},c_{{i_n}}^{dl} \ge 0,  \forall i_n \in \mathcal{I}, }\\ 
{}&{}&{\sum\limits_{i=1}^K {c_{{i_n}}^{ul}}  \leq 1,\sum\limits_{i=1}^K {c_{{i_n}}^{dl}}  \leq 1 , \forall n=1,\dots, N_c ,}\\
{}&{{\bf{C}}{\bf{.4}}}& {{\bf{Q}}_{{i_n}}^{ul} \in {\cal Q}_{{i_n}}^{ul}, {\forall i_n \in \mathcal{I}}}, \\
 {}&{}&{ \left(\mathbf{Q}_{i_{n}}^{dl}\right)_{i=1}^K \in {\cal Q}_n^{dl}, \forall n=1,\dots, N_c.}  
\end{array}
\end{array}\notag \tag{P.1}\label{P.1}
\end{equation*} 
 Constraint C.1 enforces that the latency for any MU $i_n$  to be less than or equal to the maximum tolerable delay of $T_{i_{n}}$ seconds; C.2 imposes the mentioned limit on the cloud computational resources; C.3 enforces the limited backhaul capacities in uplink and downlink; and C.4 guarantee that power budget constraint on the radio interface of both uplink and downlink is satisfied.
Note that problem (P.1) depends only on the ratios ${B_{{i_n}}^I}/{{{W^{ul}}}}$, ${B_{{i_n}}^I}/{{{C^{ul}_n}}}$, ${V_{{i_n}}}/{F_c}$, ${B_{{i_n}}^O}/{{{W^{dl}}}}$ and ${B_{{i_n}}^O}/{{{C^{dl}_n}}}$. We denote by  $\mathbf{Z}\triangleq \big( {{{\bf{Q}}^{ul}},{{\bf{Q}}^{dl}}},{{{\bf{f}}},{{\bf{c}}^{ul}}},{{{\bf{c}}^{dl}}} \big)$ the set of all optimization variables, and by $\mathcal Z$ the feasible set of (P.1). Note that (P.1) is non-convex, due to the non-convexity of the objective function and of the constraint C.1. 

\begin{remark} \normalfont
As discussed, the expression in C.1 for the latency of each user assumes that the communication and processing steps, including uplink transmission, edge-to-cloud backhaul transmission, cloud processing, cloud-to-edge backhaul transmission and downlink transmission take place one after the other for each user, as illustrated in Fig. 2. Given that the uplink, backhaul, execution and downlink latencies  may be different across the users, the communication steps of different users may not be aligned. While this fact may be exploited by a sophisticated receiver that tracks the variation of the interference power within a communication block, the resulting achievable rates are extremely difficult to characterize and optimize. In contrast, the expression of the uplink and downlink rates (3) and (8), in which interference is assumed to be caused by all users, are achievable by means of standard decoders (see, e.g., \cite{kampf2010euclidean}) and can be efficiently computed and optimized, as it will be shown in Sec. III.
\end{remark} \normalfont

\noindent \textbf{Feasibility.}
 Problem (P.1) has a non-empty feasible set if there exist matrices ${{{\bf{Q}}^{ul}}}$ and ${{{\bf{Q}}^{dl}}}$ such that the following inequalities hold \\
{a)} $T_{{i_n}} > \frac{{B_{{i_n}}^I}}{{{W^{ul}}r_{{i_n}}^{ul}\left( {{{\bf{Q}}^{ul}}} \right)}} + \frac{{B_{{i_n}}^O}}{{{W^{dl}}r_{{i_n}}^{dl}\left( {{{\bf{Q}}^{dl}}} \right)}}$, $\forall i_n \in \cal{I}$;\\
{b)} ${\sum\limits_{{i_n} \in {\cal I}} \dfrac{V_{{i_n}}/F_c}{T_{{i_n}}-\frac{{B_{{i_n}}^I}}{{{W^{ul}}r_{{i_n}}^{ul}\left( {{{\bf{Q}}^{ul}}} \right)}}  + \frac{{B_{{i_n}}^O}}{{{W^{dl}}r_{{i_n}}^{dl}\left( {{{\bf{Q}}^{dl}}} \right)}}} \leq \alpha_1 }$;\\ 
{c)} \resizebox{ 0.95\columnwidth}{!} { $\sum\limits_{i=1}^K \dfrac{B_{{i_n}}^I/C^{ul}_n}{T_{{i_n}}-\frac{{B_{{i_n}}^I}}{{{W^{ul}}r_{{i_n}}^{ul}\left( {{{\bf{Q}}^{ul}}} \right)}}+ \frac{{B_{{i_n}}^O}}{{{W^{dl}}r_{{i_n}}^{dl}\left( {{{\bf{Q}}^{dl}}} \right)}}}\leq \alpha_2$, $\forall n=1,\dots, N_c;$ }\\
{d)} \resizebox{.95 \columnwidth}{!} {$\sum\limits_{i=1}^K \dfrac{B_{{i_n}}^O/C^{dl}_n}{T_{{i_n}}-\frac{{B_{{i_n}}^I}}{{{W^{ul}}r_{{i_n}}^{ul}\left( {{{\bf{Q}}^{ul}}} \right)}}  + \frac{{B_{{i_n}}^O}}{{{W^{dl}}r_{{i_n}}^{dl}\left( {{{\bf{Q}}^{dl}}} \right)}}}\leq \alpha_3$, $\forall n=1,\dots, N_c;$}\\
\\
for some $\alpha_1, \alpha_2, \alpha_3 \geq 0$ with $\alpha_1+\alpha_2+\alpha_3 = 1$. In fact, if above conditions are met, one can always choose $f_{{i_n}}=\alpha_1^{-1}({V_{{i_n}}/F_c}){\Big(T_{{i_n}}-\frac{{B_{{i_n}}^I}}{{{W^{ul}}r_{{i_n}}^{ul}\left( {{{\bf{Q}}^{ul}}} \right)}}  - \frac{{B_{{i_n}}^O}}{{{W^{dl}}r_{{i_n}}^{dl}\left( {{{\bf{Q}}^{dl}}} \right)}}\Big)^{-1}}$, and similarly for  $c_{{i_n}}^{ul}$ and $c_{{i_n}}^{dl}$, so that conditions b), c) and d) are satisfied with equality.

\begin{remark} \normalfont \label{remark3}
When the goal is identifying the achievable trade-off curve between energy consumption and latency, assuming for simplicity that all MUs have the same latency constraint $T$, e.g., $T_{{i_n}}=T$, the following problem may also be considered
 
\begin{equation*}
\begin{array}{l}
\begin{array}{*{20}{l}}
{\mathop {{\mathop{\rm min}\nolimits} }\limits_{{{\bf{Q}}^{ul}},{{\bf{Q}}^{dl}},{\bf{f}},{{\bf{c}}^{ul}},{{\bf{c}}^{dl}},T} }&{E\left( {{{\bf{Q}}^{ul}}},{{{\bf{Q}}^{dl}}} \right)}+ \lambda T\\
\end{array}\\
\begin{array}{*{20}{l}}
\;\;\;\;\;\;\;\;\;\; \text{s.t.}&{{\bf{C}}.\mathbf{1}}\;\;\;\;{\frac{{B_{{i_n}}^I}}{{{W^{ul}}r_{{i_n}}^{ul}\left( {{{\bf{Q}}^{ul}}} \right)}} + \frac{{B_{{i_n}}^I}}{{c_{{i_n}}^{ul}{C^{ul}_n}}} + \frac{{{V_{{i_n}}}}}{{{f_{{i_n}}}{F_c}}}}\\
{}&{}\;\;\;\;\;\;\;\;\;\;{ +\frac{{B_{{i_n}}^O}}{{c_{{i_n}}^{dl}{C^{dl}_n}}}  + \frac{{B_{{i_n}}^O}}{{{W^{dl}}r_{{i_n}}^{dl}\left( {{{\bf{Q}}^{dl}}} \right)}} \le {T}}, \forall i_n \in \mathcal{I}, \\
{}&{\bf{C}}.\mathbf{2}-{\bf{C}}.\mathbf{4} \;\; \text{of (P.1)},  \\
\end{array}
\end{array}\notag \tag{P.2}\label{P.2}
\end{equation*} 
where $\lambda > 0$ is a parameter identifying the desired relative weight between energy and latency minimization.
\label{remark3}
\end{remark}
\begin{remark} \normalfont
The problem formulation (P.1) can be easily extended to account for a more general backhaul topology in which the ceNBs are connected to the cloud via a multi-hop network with predefined routes between ceNBs and cloud. We do not further elaborate on this model here, because of the space limitation. 

\label{remark4}
\end{remark}

\begin{remark} \normalfont
Problem (P.1) was tacked in \cite{sar02} for the special case where the joint optimization was over radio and computational resources and only in the uplink direction. Problem (P.1), instead, considers a general setup in which joint optimization carried over backhaul capacities in both uplink and downlink, as well as the optimization over the downlink radio transmission from the ceNB to the MU. This generalization entails different formulations for the objective function and for the latency constraint from that in \cite{sar02} and thus calls for novel SCA-based solution methods.
\end{remark}

\section{successive convex approximation optimization} \label{sec:sca}
Problem (P.1) is non-convex due to the non-convexity of the objective function and of the constraint C.1. To address this issue, we leverage the SCA method proposed in \cite{scutari2014distributed, scutari2016parallel} for general non-convex problems. The SCA algorithm in \cite{scutari2014distributed, scutari2016parallel} is proved to converge to a stationaty solution of the  (NP-hard) non-convex probem by solving a sequence of convex sub-problems, each one of which can be solved in polynomial time, e.g., by interior-point methods \cite{nesterov1994interior}. To do so, we need to identify convex approximants for the objective function and for the non-convex constraints C.1 that satisfy the conditions specified in \cite{scutari2014distributed, scutari2016parallel} as discussed next.

\subsection{Convex Surrogate for the Objective Function} \label{con_obj}
Define as $\cal{K}\supseteq \cal{Z}$ any compact convex set containing the   feasible set $\cal{Z}$ such that all functions in (P.1) are well defined on it. Note that such a set always exists. This set exists by the same arguments used in \cite[Sec. IV-B]{sar02}. Let ${\bf{Z}}\left( v \right) \triangleq \big( {{{\bf{Q}}^{ul}}\left( v \right),{{\bf{Q}}^{dl}}\left( v \right)}, {{{\bf{f}}}\left( v \right),{{\bf{c}}^{ul}}\left( v \right)},{{{\bf{c}}^{dl}}\left( v \right)} \big)$, with $v$ being the current iterate index of the SCA algorithm. According to \cite{scutari2014distributed, scutari2016parallel}, in order to be used within the SCA scheme, a convex approximant $\tilde E\left( {{{\bf{Z}}};{\bf{Z}}\left( v \right)} \right)$ of the objective function $ E\left( {{{\bf{Q}}^{ul}}}, {{{\bf{Q}}^{dl}}} \right)$ around the current feasible iterate ${\bf{Z}}\left( v \right) \in \cal{Z}$ must satisfy the following properties (\!\!\cite[Sec. II]{scutari2014distributed}):\\
A1: $\tilde E\left( {\bullet;{{\bf{Z}}}\left( v \right)} \right)$ is uniformly strongly convex on $\cal{K}$;\\
A2: ${\nabla _{{{\bf{Q}}^{u{l^*}}}}}\tilde E\left( {{{\bf{Z}}}\left( v \right);{{\bf{Z}}}\left( v \right)} \right) = {\nabla _{{{\bf{Q}}^{u{l^*}}}}}E\left( {{\bf{Q}}^{ul}}\left( v \right), {{\bf{Q}}^{dl}}\left( v \right) \right)$ and ${\nabla _{{{\bf{Q}}^{d{l^*}}}}}\tilde E\left( {{{\bf{Z}}}\left( v \right);{{\bf{Z}}}\left( v \right)} \right) = {\nabla _{{{\bf{Q}}^{d{l^*}}}}}E\left({{\bf{Q}}^{ul}}\left( v \right), {{\bf{Q}}^{dl}}\left( v \right) \right)$, for all ${{\bf{Z}}}\left( v \right) \in \cal{Z}$;\\
A3: ${\nabla _{{{\bf{Z}}^{{*}}}}}\tilde E\left(\bullet;\bullet \right) $ is Lipschitz continuous on $\cal{K} \times \cal{Z}$; \\
where  ${\nabla _{{{\bf{x}}^{{*}}}}}f(\bf x;y)$ denotes the conjugate gradient of the function $f(\bf x;y)$ with respect to its first argument $\bf{x}$. We note that, besides the convexity and smoothness conditions A1 and A3, A2 enforces that the first-order behavior of the approximant be the same as for the original function. In what follows, we derive an approximant for the objective function (14) satisfying  A1-A3 above. Let us write
\begin{equation} \label{E_apprx_dl}
\begin{split}
& \tilde E\left( {{{\bf{Z}}}}; {{\bf{Z}}}\left( v \right) \right)    \triangleq \sum\limits_{{i_n} \in {\cal I}} {{{\tilde E}_{{i_n}}}\left( {{{\bf{Z}}_{{i_n}}}}; {{\bf{Z}}}\left( v \right) \right)+ {\rm \bar E}\left( {{{\bf{Z}}}}; {{\bf{Z}}}\left( v \right) \right)   } \\
= &  \sum\limits_{{i_n} \in {\cal I}} \Big( {\tilde E_{{i_n}}^{ul}}\left( {{{\bf{Q}}^{ul}}};{{{\bf{Q}}^{ul}}}(v) \right) +{\tilde E_{{i_n}}^{dl}}\left( {{{\bf{Q}}^{dl}}};{{{\bf{Q}}^{dl}}}(v) \right) \Big)  \\
& + {\rm \bar E}\left( {{{\bf{Z}}}}; {{\bf{Z}}}\left( v \right) \right) ,
\end{split}
\end{equation}where ${\bf{Z}}_{{i_n}} \triangleq \big( {{{\bf{Q}}^{ul}_{{i_n}}},{{\bf{Q}}^{dl}_{{i_n}}}},{{{{f}}}_{{i_n}},{{{c}}^{ul}_{{i_n}}}},{{{{c}}^{dl}_{{i_n}}}} \big)$ with the function ${\rm \bar E}\left( {{{\bf{Z}}}}; {{\bf{Z}}}\left( v \right) \right) \triangleq \frac{{{{{\gamma_{{q^{ul}}}}}}}}{2}\sum_{{i_n} \in {\cal I}}{\left\| {{{\bf{Q}}^{ul}} - {{\bf{Q}}^{ul}}\left( v \right)} \right\|^2}+\frac{{{{{\gamma_{{q^{dl}}}}}}}}{2}\sum_{{i_n} \in {\cal I}}{\left\| {{{\bf{Q}}^{dl}} - {{\bf{Q}}^{dl}}\left( v \right)} \right\|^2} + \frac{{{\gamma_f}}}{2}{\left\| {{\bf{f}} - {\bf{f}}\left( v \right)} \right\|^2} + \frac{{{\gamma_{{c^{ul}}}}}}{2}{\left\| {{{\bf{c}}^{ul}} - {{\bf{c}}^{ul}}\left( v \right)} \right\|^2} + \frac{{{\gamma_{{c^{dl}}}}}}{2}{\left\| {{{\bf{c}}^{dl}} - {{\bf{c}}^{dl}}\left( v \right)} \right\|^2}$  is added to make the approximant objective function (\ref{E_apprx_dl}) strongly convex on $\cal{K}$, with ${\gamma_{q^{ul}}},{\gamma_{q^{dl}}}, {\gamma_f}, {\gamma_{{c^{ul}}}}$ and ${\gamma_{{c^{dl}}}}$ being arbitrary positive constants (see \cite{scutari2014distributed, scutari2016parallel}{}).The functions ${\tilde E_{{i_n}}^{ul}}\left( {{{\bf{Q}}^{ul}}};{{{\bf{Q}}^{ul}}}(v) \right)$ and ${\tilde E_{{i_n}}^{dl}}\left( {{{\bf{Q}}^{dl}}};{{{\bf{Q}}^{dl}}}(v) \right)$
are the  convex approximants for the uplink and downlink energy terms, respectively, as derived next.

\textit{1) Convex approximant for} $E_{{i_n}}^{ul} \left( {{{\bf{Q}}^{ul}}}\right)$: It is not difficult to check that an approximant that utilize the ``partial'' convexity of the uplink energy function (\ref{off_ene}) can be obtained as (cf. \cite[Sec. IV-B]{sar02})
\begin{equation} \label{Ein_apprx} 
\begin{split}
{\tilde E_{{i_n}}^{ul}}&\left( {{{\bf{Q}}^{ul}}};{{{\bf{Q}}^{ul}}}(v) \right)      =  \;      \text{tr}\left( {{\bf{Q}}_{{i_n}}^{ul}\left( v \right)} \right)\frac{{B_{{i_n}}^I}}{ {r_{{i_n}}^{ul}\left( {{\bf{Q}}_{{i_n}}^{ul},{\bf{Q}}_{-i_{n}}^{ul}\left( v \right)} \right)}} \\
& + \text{tr}\left( {{\bf{Q}}_{{i_n}}^{ul}} \right)\frac{{B_{{i_n}}^I}}{{r_{{i_n}}^{ul}\left( {{\bf{Q}}_{{i_n}}^{ul}\left( v \right),{\bf{Q}}_{-i_{n}}^{ul}\left( v \right)} \right)}}\\
&  + \sum\limits_{{j_m} \in {\cal I},m \ne n} {\left\langle {{\nabla _{{\bf{Q}}{{_{{i_n}}^{ul}}^*}}}{E_{{j_m}}}\left( {{{\bf{Q}}^{ul}}\left( v \right)} \right),{\bf{Q}}_{{i_n}}^{ul} - {\bf{Q}}_{{i_n}}^{ul}\left( v \right)} \right\rangle }, 
\end{split}
\end{equation}
where $\left\langle {{\bf{A}},{\bf{B}}} \right\rangle  \triangleq {\mathop{\rm Re}\nolimits} \left\{ {\text{tr}\left( {{{\bf{A}}^H}{\bf{B}}} \right)} \right\}$, and the conjugate gradient ${{\nabla _{{\bf{Q}}{{_{{i_n}}^{ul}}^*}}}{E_{{j_m}}}\left( {{{\bf{Q}}^{ul}}\left( v \right)} \right)}$ is given by \cite[eq. (18)]{sar02}
\begin{equation} \label{Ein_gradient} 
\begin{split}
 {\nabla _{{\bf{Q}}{{_{{i_n}}^{ul}}^*}}}&{E_{{j_m}}}\left( {{{\bf{Q}}^{ul}}\left( v \right)} \right) = \;  \frac{{\text{tr}\left( {{\bf{Q}}_{{j_m}}^{ul}\left( v \right)} \right)\Delta _{{j_m}}^{ul}\left( {{{\bf{Q}}^{ul}}\left( v \right)} \right)}}{{\log \left( 2 \right){r_{{j_m}}^{ul}}\left( {{{\bf{Q}}^{ul}}\left( v \right)} \right)}}\\
& \cdot[{\bf{H}}_{{i_n}m}^H({\bf{R}}_m^{ul}{\left( {{\bf{Q}}_{ - {j_m}}^{ul}\left( v \right)} \right)^{ - 1}} - ({\bf{R}}_m^{ul}\left( {{\bf{Q}}_{ - {j_m}}^{ul}\left( v \right)} \right) \\
& + {{\bf{H}}_{{j_m}}}{\bf{Q}}_{{j_m}}^{ul}\left( v \right){\bf{H}}_{{j_m}}^H{)^{ - 1}}){{\bf{H}}_{{i_n}m}} ].
\end{split}
\end{equation}

\textit{2) Convex approximant for} $E_{{i_n}}^{dl} \left( {{{\bf{Q}}^{dl}}}\right)$:
To obtain the desired convex approximant of downlink energy in (\ref{dl_ene}), we need to construct the convex surrogate for the downlink rate function, i.e., ${{\tilde r_{{i_n}}^{dl}\left( {{{\bf{Q}}^{dl}};{{\bf{Q}}^{dl}}\left( v \right)} \right)}}$. To obtain such an approximant, we exploit first the concave-convex structure of the rate functions $r_{{i_n}}^{dl}\left( {{{\bf{Q}}^{dl}}} \right) $
\begin{equation} \label{eq_6} 
\begin{split}
r_{{i_n}}^{dl}\left( {{{\bf{Q}}^{dl}}} \right)
 = \; & \underbrace {{{\log }_2}\det \left( {{\bf{R}}_n^{dl}\left( {{\bf{Q}}_{ - {i_n}}^{dl}} \right) + {\bf{H}}_{{i_n}}^H{{\bf{H}}_{{i_n}}}{\bf{Q}}_{{i_n}}^{dl}} \right)}_{r{{_{{i_n}}^{dl}}^ + }\left( {{{\bf{Q}}^{dl}}} \right)} \\
& - \underbrace {{{\log }_2}\det \left( {{\bf{R}}_n^{dl}\left( {{\bf{Q}}_{ - {i_n}}^{dl}} \right)} \right)}_{r{{_{{i_n}}^{dl}}^ - }\left( {{\bf{Q}}_{-i_{n}}^{dl}} \right)},
\end{split}
\end{equation} 
where ${r{{_{{i_n}}^{dl}}^ + }\left( {{{\bf{Q}}^{dl}}} \right)}$ and ${r{{_{{i_n}}^{dl}}^ - }\left( {{{\bf{Q}}^{dl}_{-i_{n}}}} \right)}$ are concave functions. The convex approximant is then obtained as
\begin{equation} \label{g_t_apprx_2} 
\begin{split}
\tilde r_{{i_n}}^{dl}&\left( {{{\bf{Q}}^{dl}};{{\bf{Q}}^{dl}}\left( v \right)} \right) =\; r_{{i_n}}^{dl + }\left( {{{\bf{Q}}^{dl}}} \right) - r_{{i_n}}^{dl - }\left( {{\bf{Q}}_{-i_{n}}^{dl}\left( v \right)} \right) \\
&-  {\sum\limits_{j_m \in \cal{I} } {\left\langle {{\nabla _{{\bf{Q}}{{_{{j_m}}^{ul}}^*}}}r{{_{{i_n}}^{dl}}^ - }\left( {{\bf{Q}}_{{-i_n}}^{dl}\left( v \right)} \right),{\bf{Q}}_{{j_m}}^{dl} - {\bf{Q}}_{{j_m}}^{dl}\left( v \right)} \right\rangle } }  ,
\end{split}
\end{equation}
with 
\begin{equation} \label{g_t_apprx_4} 
{\nabla _{{\bf{Q}}{{_{{j_m}}^{dl}}^*}}}r_{{i_n}}^{dl - }\left( {{\bf{Q}}_{{-i_n}}^{dl}\left( v \right)} \right) = {\bf{H}}_{{j_m}n}^H{\bf{R}}_n^{dl}{\left( {{\bf{Q}}_{-i_{n}}^{dl}\left( v \right)} \right)^{ - 1}}{{\bf{H}}_{{j_m}n}}.
\end{equation}
Then, we can simply obtain the convex approximant for $E_{{i_n}}^{dl} \left( {{{\bf{Q}}^{dl}}}\right)$ as
\begin{equation}
\tilde E_{{i_n}}^{dl} \left( {{{\bf{Q}}^{dl}};{{\bf{Q}}^{dl}}\left( v \right)} \right) ={B_{{i_n}}^O}\frac{{d_{{i_n}}}}{{\tilde r_{{i_n}}^{dl}\left( {{{\bf{Q}}^{dl}};{{\bf{Q}}^{dl}}\left( v \right)} \right)}}.
\end{equation}
\begin{remark} \normalfont
The key advantage of SCA \cite{scutari2014distributed, scutari2016parallel} as compared to more conventional approaches such as Difference-of-Convex (DC) programming \cite{dc} is that the convex surrogate $\tilde E\left( {{{\bf{Z}}}}; {{\bf{Z}}}\left( v \right) \right)$ need not be a global upper bound on the function $E({\bf Q}^{ul}, {\bf Q}^{dl})$ -- a condition that appears to be difficult to ensure for the objective function of (P.1).
\end{remark}

%

%
\subsection{Inner Convexification of the Constraints} \label{con_con}
%
Next, let us consider the latency constraint C.1. Note that this constraint differs from the corresponding latency constraint in \cite{sar02} by virtue of the contributions due to downlink and backhaul transmissions. Let us define the non-convex part of the left-hand side of C.1 as
\begin{equation} \label{g_t} 
{g_{{i_n}}}\left( {{{\bf{Q}}^{ul}},{{\bf{Q}}^{dl}}} \right) \triangleq \frac{{B_{{i_n}}^I}}{{W^{ul}}{r_{{i_n}}^{ul}\left( {{{\bf{Q}}^{ul}}} \right)}} + \frac{{B_{{i_n}}^O}}{{W^{dl}}{r_{{i_n}}^{dl}\left( {{{\bf{Q}}^{dl}}} \right)}}.
\end{equation}
To apply SCA, we need to obtain an approximant $\tilde g_{{i_n}}\big({\bf{Q}}^{ul},{\bf{Q}}^{dl};\textbf{Z}(v)\big)$ at the current iterate ${\bf{Z}}\left( v \right) ) \in \cal{Z}$ that satisfies the following properties (\!\! \cite[Sec. II]{scutari2014distributed}):\\
B1: $\tilde g_{{i_n}}\big(\bullet,\textbf{Z}(v)\big)$ is uniformly convex on $\mathcal{K}$;\\
B2: ${\nabla _{{{\bf{Z}}^{*}}}}\tilde g_{{i_n}}\big( {\bf{Q}}^{ul}\left( v \right),{\bf{Q}}^{dl}\left( v \right);{\textbf{Z}(v)} \big) = {\nabla _{{{\bf{Z}}^{*}}}}g_{{i_n}}\big( {\bf{Q}}^{ul}\left( v \right),{\bf{Q}}^{dl}\left( v \right) \big)$, for all $\textbf{Z}(v) \in \mathcal{Z}$;\\
B3: ${\nabla _{{{\bf{Z}}^{*}}}}\tilde g_{{i_n}}(\bullet;\bullet)$ is continuous on $\mathcal{K}\times \mathcal{Z}$; \\
B4: $\tilde g_{{i_n}}\big({\bf{Q}}^{ul},{\bf{Q}}^{dl};\textbf{Z}(v)\big)\geq  g_{{i_n}}\big({\bf{Q}}^{ul},{\bf{Q}}^{dl}\big)$, for all $\big({\bf{Q}}^{ul},{\bf{Q}}^{dl}\big) \in \cal{K}$ and $\textbf{Z}(v) \in \mathcal{Z}$; \\
B5: $\tilde g_{{i_n}}\big({\bf{Q}}^{ul}\left( v \right),{\bf{Q}}^{dl}\left( v \right);\textbf{Z}(v)\big)= g_{{i_n}}\big({\bf{Q}}^{ul}\left( v \right),{\bf{Q}}^{dl}\left( v \right)\big)$, for all $\textbf{Z}(v) \in \mathcal{Z}$;\\
B6: $\tilde g_{{i_n}}(\bullet;\bullet)$ is Lipschitz continuous on $\mathcal{K}\times \mathcal{Z}$. \\
Besides the first-order behavior and smoothness conditions B2, B3, B6, the key assumptions B1, B4 and B5 enforce that the approximant $\tilde g_{{i_n}}\big({\bf{Q}}^{ul},{\bf{Q}}^{dl};\textbf{Z}(v)\big)$ be a locally tight (condition B5)   convex (condition B1) upper bound (condition B4) on the original constraint ${g_{{i_n}}}\left( {{{\bf{Q}}^{ul}},{{\bf{Q}}^{dl}}} \right)$. The desired surrogate approximation ${\tilde g}_{{i_n}}\left( {{{\bf{Q}}^{ul}},{{\bf{Q}}^{dl}}; \textbf{Z}(v) }  \right)$ is then obtained from (\ref{g_t_apprx_2}) as
\begin{equation} \label{g_t_apprx} 
\begin{split}
 {{\tilde g}_{{i_n}}}   \left( {{{\bf{Q}}^{ul}},{{\bf{Q}}^{dl}}; \textbf{Z}(v)} \right) &  \triangleq \;  \frac{{B_{{i_n}}^I}}{{W^{ul}}{\tilde r_{{i_n}}^{ul}\left( {{{\bf{Q}}^{ul}};{{\bf{Q}}^{ul}}\left( v \right)} \right)}} \\
 &+  \frac{{B_{{i_n}}^O}}{{W^{dl}}{\tilde r_{{i_n}}^{dl}\left( {{{\bf{Q}}^{dl}};{{\bf{Q}}^{dl}}\left( v \right)} \right)}},
\end{split}
\end{equation}

The proposed approximant  (\ref{g_t_apprx}) is an upper bound (condition B4) and is a convex function of $ {\bf{Q}}^{ul} $ and ${\bf{Q}}^{dl}$ (condition B1). This is because both terms in (\ref{g_t_apprx}) are the reciprocal of a concave and positive function, and the sum of the two convex functions is convex. Furthermore, it is easy to show that the original constraint (\ref{g_t}) and its convex approximant (\ref{g_t_apprx}) have the same first-order behavior (condition B2) by evaluating the gradient of both functions at the current iterate. The remaining properties B3-B6 can also be checked in a similar manner to \cite[Sec. IV-B]{sar02}. For instance, since the approximant functions (\ref{Ein_apprx}) and (\ref{g_t_apprx_2}) are twice continuously differentiable over the compact convex set $\cal{K}\supseteq \cal{Z}$, the Lipschitz continuity of their conjugate gradients follows readily.

%
\subsection{SCA Algorithm}
The SCA algorithm operates by iteratively solving the following problem around the current iterate ${\bf{Z}}\left( v \right) \in \cal{Z}$,
\begin{align*}
& {\bf{\hat Z}}\left( {{\bf{Z}}\left( v \right)} \right) \triangleq \underset{\mathbf{Q}^{ul},\mathbf{Q}^{dl},\mathbf{f},\mathbf{c}^{ul},\mathbf{c}^{dl}}{\text{argmin}}
   \begin{aligned}[t]
      &{\tilde E\left( {{{\bf{Z}}}}; {{\bf{Z}}}\left( v \right) \right)   } \\
   \end{aligned} \notag \\
   &\;\;\;\;\;\;\;\;\;\;\;\;\;\;\;\;\;\;\;\;\;\;\;\;\;\;\;\;\; \text{s.t.}  \\
 & \begin{aligned}[t]
      & \mathbf{C.1}  \;\;\;\; {{\tilde g}_{{i_n}}}\left( {{{\bf{Q}}^{ul}},{{\bf{Q}}^{dl}}; \textbf{Z}(v) } \right) + \frac{{B_{{i_n}}^I}}{{c_{{i_n}}^{ul}{C^{ul}_n}}} + \frac{{B_{{i_n}}^O}}{{c_{{i_n}}^{dl}{C^{dl}_n}}}\\
      & \;\;\;\;\;\;\;\;\; + \frac{{{V_{{i_n}}}}}{{{f_{{i_n}}}{F_c}}}  \le {T_{{i_n}}}, \forall i_n \in \cal{I},
   \end{aligned}  \\
   & {}{\bf{C}}.\mathbf{2}-{\bf{C}}.\mathbf{4} \;\; \text{of (P.1)}. \tag{P.3}\label{P.3}  
\end{align*} 
 The unique solution of the strongly convex optimization problem (P.3) is denoted $
{\bf{\hat Z}}\big( {{\bf{Z}}\left( v \right)} \big) \triangleq \big({{{\bf{\hat Q}}}^{ul}},{{{\bf{\hat Q}}}^{dl}},{\bf{\hat f}},{{{\bf{\hat c}}}^{ul}}
,{{{\bf{\hat c}}}^{dl}}\big)$. Note that ${\tilde E\left( {{{\bf{Z}}}}; {{\bf{Z}}}\left( v \right) \right)   }$ is a function of $\bf{Z}$, given the current iterate ${{\bf{Z}}}\left( v \right)$. 

The SCA scheme is summarized in Algorithm 1. In step 2, the termination criterion is $\left| {E\left( {{{\bf{Q}}^{ul}}\left( {v + 1} \right)}, {{{\bf{Q}}^{dl}}\left( {v + 1} \right)} \right) - E\left( {{{\bf{Q}}^{ul}}\left( v \right)}, {{{\bf{Q}}^{dl}}\left( v \right)} \right)} \right| \le \delta $, where $\delta  > 0$ is the desired accuracy. The step size rule we used is $\gamma \left( v \right) = \gamma \left( {v - 1} \right)\left( {1 - \alpha \gamma \left( {v - 1} \right)} \right)$ with $\gamma \left( 0 \right) \in \left( {0,1} \right]$ and $\alpha  \in \left( {0,1/\gamma \left( 0 \right)} \right)$ (other step size rules can also be adopted, see\cite{scutari2014distributed, scutari2016parallel}). Algorithm 1 converges to a stationary point of the problem (\ref{P.1}) in the sense of \cite[Theorem 2]{sar02}.

\begin{remark} \normalfont
The SCA scheme can also be easily adapted to tackle the weighted sum problem (\ref{P.2}) discussed in Remark \ref{remark3}. This alternative formulation has the key advantage that the identification of an initial feasible point ${\bf{Z}}\left( 0 \right) \triangleq \big( {{\bf{Q}}^{ul}}\left( 0 \right),{{\bf{Q}}^{dl}}\left( 0 \right),{{\bf{f}}}\left( 0 \right),$ ${{\bf{c}}^{ul}}\left( 0 \right),{{\bf{c}}^{dl}}\left( 0 \right), {{{{T}}}\left( 0 \right)} \big)$ for the SCA is a trivial task. This is because one can always select a value of $T(0)$ that satisfies the constraint C.1 in (P.2) for given values of the other variables.
\end{remark}
\begin{remark}\normalfont\
Each instance of the optimization problem (P.3) tackled by SCA can be solved with complexity $O(\text{max}\{n^3, n^2m\})$ using interior-points methods \cite[Ch. 1]{boyd2004convex}, where $n$ is the size of the optimization variables, namely $\left(2N_{T_{i_n}}^2+3\right) KN_c$, and $m$ is the number of constraints, namely $m=7KN_c+3N_c+1$. Note that the complexity scales polynomially with the number $K$ of users and with all system parameters. While here we have focused on a centralized implementation, the complexity could be further reduced by developing distributed solutions as described in \cite[Sec. IV]{scutari2014distributed}. Finally, we would also like to mention that, in practice, rather than solving  problem (P.3) using SCA at each time slot for the given realization of the channels, it would be possible to solve the problem for a number of representative channels so as to build a sufficiently dense look-up table. More interestingly, as recently explored in \cite{sunlearning} for power allocation in an interference channel, one could use such representative channels to train a neural network, or another learning machine, to ``interpolate'' the solution to other channel realizations. We leave these aspects for future investigations.
\end{remark}

\begin{algorithm}
\caption{SCA Solution for (P.3) }\label{alg:SCA_Alg}
\begin{algorithmic}[1]
    \INPUT Parameters from Table I; ${\bf{Z}}\left( 0 \right) \in {\cal Z}$; $v = 0$; ${\left\{ {\gamma \left( v \right)} \right\}_v} \in \left( {0,1} \right]$;   ${\gamma_{q^{ul}}},{\gamma_{q^{dl}}},{\gamma_{f}},{\gamma_{c^{ul}}},{\gamma_{c^{dl}}} > 0$.
    \State If ${\bf{Z}}\left( v \right)$ satisfies the termination criterion, {stop}.
    \State Compute ${\bf{\hat Z}}\left( {{\bf{Z}}\left( v \right)} \right)$ from (\ref{P.3}).
    \State Set ${\bf{Z}}\left( {v + 1} \right) = {\bf{Z}}\left( v \right) + \gamma \left( v \right)\left( {{\bf{\hat Z}}\left( {{\bf{Z}}\left( v \right)} \right) - {\bf{Z}}\left( v \right)} \right)$.
    \State $v \leftarrow v + 1$, and return to step $1$.
    \OUTPUT $\mathbf{Z}= \big( {{{\bf{Q}}^{ul}},{{\bf{Q}}^{dl}}},{{{\bf{f}}},{{\bf{c}}^{ul}}},{{{\bf{c}}^{dl}}} \big)$.
\end{algorithmic}
\end{algorithm}

%
%
%
%
\section{USER SELECTION} 
\label{selection}
In the previous section, we assumed that a given number of active users, namely $K$ per cell, was scheduled for transmission. The premise of this section, is that, if too many MUs simultaneously choose to offload their computational tasks, the resulting interference on the wireless channel may require an energy consumption at the mobile for wireless transmission that exceeds the energy that would be needed for local computing at some MUs. Moreover, the backhaul and computing delays may make the latency constraint in problem (P.1) impossible to satisfy, and thus problem (P.1) infeasible. For theses reasons, in this section, we consider user selection with the aim of maximizing the number of MUs that perform offloading while guaranteeing that the selected MUs can satisfy their latency constraints and, at the same time, consume less energy than with local computing. In the rest of this section, the local computation energy model is first elaborated on and then the user scheduling problem is formulated and tackled by integrating SCA with smooth $l_p$-norm approximation methods.

%
\subsection{Local Computation Energy}
When the application is executed at the mobile device, the
energy consumption $E_{i_{n}}^M$ is determined by the number of CPU cycles required by the application, $V_{i_{n}}$, and by the clock frequency of the device chip, which is denoted here as $F_{i_{n}}$. In particular, for CMOS circuits, the energy per operation is proportional to the square of the supply voltage to the chip, and when the supply voltage is low, the clock frequency of the chip is a linear function of the voltage supply \cite{cpu}. As a result, the mobile energy for computing can be expressed as
\begin{equation} \label{ec}
E_{i_{n}}^M=\kappa V_{i_{n}} F_{i_{n}}^2 ,
\end{equation}
where $\kappa$ is the effective switched capacitance, which depends on the MU processor architecture, and the clock frequency is selected so as to meet the latency constraint, yielding 
\begin{equation}
F_{i_{n}} = \frac{V_{i_{n}}}{T_{i_{n}}}.
\end{equation}
By plugging this into (\ref{ec}), we obtain the total consumption energy for mobile execution as
\begin{equation} \label{local}
E_{i_{n}}^M = \kappa \frac{V_{i_{n}}^3}{T_{i_{n}}^2}.
\end{equation}
For each MU $i_n$, offloading is advantageous when the energy for local mobile computing is higher than the energy required for offloading, i.e.,
  \begin{equation} \label{condition}
E_{i_{n}}^M \geq {{E_{{i_n}}^{ul}}\left( {{\bf{Q}}^{ul}} \right)},
\end{equation}
where ${{E_{{i_n}}^{ul}}\left( {{\bf{Q}}^{ul}} \right)}$ is given in (6). Note that here we consider for brevity only the uplink energy contribution in (\ref{mu-ene}).

\subsection{User Scheduling}

To proceed, we introduce the auxiliary slack variables $(x_{{i_n}},y_{{i_n}})$ for each MU $i_n$ measuring the violation of the latency constraint C.1 in (\ref{P.1}) and the energy constraint (\ref{condition}), respectively. Our system design becomes maximizing the number of MUs that can perform offloading, while satisfying the latency constraints and guaranteeing energy savings with respect to local computation when offloading is performed. This amounts to maximizing the number of MUs $i_n$ with no violation of the mentioned constraints, i.e., with $x_{{i_n}} = 0$ and $y_{{i_n}} = 0$. This is done here by minimizing the $\ell_0$-norm
\begin{equation} \label{lo}
\|\mathbf{x}\|_0+\|\mathbf{y}\|_0 =\sum_{i_n\in \mathcal{X}} \mathbf{I}(x_{{i_n}} > 0) + \sum_{i_n\in \mathcal{X}} \mathbf{I}(y_{{i_n}} > 0),
\end{equation}
where $\mathbf{I}$ is an indicator function that returns $1$ for $x_{{i_n}},y_{{i_n}} > 0$ and $0$ otherwise, also define $\mathbf{x}\triangleq [x_{{i_n}}]_{i_n \in \mathcal{X}}$ and $\mathbf{y}\triangleq [y_{{i_n}}]_{{i_n} \in \mathcal{X}}$. The sum of the $l_0$-norms of the slack vectors $\mathbf x$ and $\mathbf y$  in (\ref{lo}) counts the number of constraints C.1 and C.2 that are violated by the users.  Therefore, minimizing this sum enforces the selection of users that satisfy the largest number of constraints. Accordingly, the problem, defined for generality over any arbitrary subset of users $\mathcal{X} \subseteq \mathcal{I}$, reads
\begin{equation*}
\begin{array}{lcl}
\begin{array}{*{20}{l}}
{\mathop {{\mathop{\rm min}\nolimits} }\limits_{{{\bf{Q}}^{ul}},{{\bf{Q}}^{dl}},{\bf{f}},{{\bf{c}}^{ul}},{{\bf{c}}^{dl}}, \bf{x}, \bf{y} } }&{\|\mathbf{x}\|}_0 + {\|\mathbf{y}\|}_0\\
\end{array}\\
\begin{array}{*{20}{l}}
 \;\;\;\;\;\;\;\;\;\;\;  \text{s.t.}& {{\bf{C}}.\mathbf{1}}\;\;\; {\frac{{B_{{i_n}}^I}}{{{W^{ul}}r_{{i_n}}^{ul}\left( {{{\bf{Q}}^{ul}}} \right)}} + \frac{{B_{{i_n}}^I}}{{c_{{i_n}}^{ul}{C^{ul}_n}}} + \frac{{{V_{{i_n}}}}}{{{f_{{i_n}}}{F_c}}}}+ \frac{{B_{{i_n}}^O}}{{c_{{i_n}}^{dl}{C^{dl}_n}}}\\
{}&{}{\;\;\;\;\;\;\;\;\;  + \frac{{B_{{i_n}}^O}}{{{W^{dl}}r_{{i_n}}^{dl}\left( {{{\bf{Q}}^{dl}}} \right)}}- {T_{{i_n}}}\le x_{{i_n}} }, \forall i_n \in \mathcal{X},\\
{}&{{\bf{C}}{\bf{.2}}}\;\;\;\;\;{{E_{{i_n}}^{ul}}\left( {{\bf{Q}}^{ul}} \right)} - E_{i_{n}}^M \le y_{{i_n}}, \forall i_n \in \mathcal{X}, \\
{}&{{\bf{C}}{\bf{.3}}}\;\;\;\;\;x_{i_{n}} \geq 0, y_{i_{n}} \geq 0, \forall i_n \in \mathcal{X},  \\
& {}{\bf{C}}.\mathbf{2}-{\bf{C}}.\mathbf{4} \;\; \text{of (P.1)}, \forall i_n \in \mathcal{X}.
\end{array}
\end{array}\notag \tag{P.4}\label{P.4}
\end{equation*} 
Define ${\bf{Z}} \triangleq \big( {{{\bf{Q}}^{ul}},{{\bf{Q}}^{dl}}},{{{\bf{f}}},{{\bf{c}}^{ul}}},{{{\bf{c}}^{dl}}}, \bf{x}, \bf{y} \big)$.

The scheduling algorithm is described in Algorithm 2. The algorithm maximizes the number of scheduled MUs by progressively removing the MUs $i_n$ that have the largest entries in the vector $\mathbf{w}\triangleq\mathbf{x}/\sum_{{i_{n}\in\mathcal{X}}}{x_{i_{n}}}+\mathbf{y}/\sum_{{i_{n}\in\mathcal{X}}}{y_{i_{n}}}$, which measures the relative amount, with respect to all users in $\mathcal{X}$, by which a user violates the two constraints. Note that the normalizations by $\sum_{{i_{n}\in\mathcal{X}}}x_{i_{n}}$ and $\sum_{{i_{n}\in\mathcal{X}}} y_{i_{n}}$ ensure that the two constraints are considered on an equal footing. The outlined iterative procedure is repeated until a subset of users $\mathcal{X}^{\ast}$ is found for which problem (P.4) returns vector $\mathbf{w}$ that is close to zero,  signifying feasibility of offloading under constraints C.1-C.4 of (\ref{P.1}) as well as (\ref{condition}). We observe that, unlike the admission control scheme  in \cite[Algorithm 2]{ran}, the proposed algorithm requires two set of auxiliary variables in order to account for the constraints C.1 and C.2.

Algorithm 2 returns the solution $s^{\ast}$, from which the set of MUs scheduled for offloading is obtained as $\mathcal{X}^{\ast} \triangleq \left\{ \pi_{1}, \ldots, {{\pi_{{s}^{\ast}}}} \right\}$. In more details, upon obtaining the solution
of (P.4), the set of MUs is ordered according to the respective
values of the entries of vector $\mathbf{w}$. Then, the subset $\mathcal{X}^{\ast}$ of scheduled users is computed by bisection. In particular, bisection searches for the minimum number of users in the interval $[0, KN_c]$, where $KN_c$ is the total number of users, that should be removed, so that the rest
of the users can be scheduled for offloading while satisfying
the desired constraints. Specifically, as described in Algorithm \ref{alg:2}, set $\mathcal{X}^{\ast}$ is defined as $\mathcal{X}^{\ast} \triangleq \left\{ \pi_{1}, \ldots, {{\pi_{{s}^{\ast}}}} \right\}$, where the value of $s^{\ast} \in [0,KN_c]$ is found by successively searching within the interval $\left[K_{\text{low}},K_{\text{up}}\right]$, which is initialized as $\left[0,KN_c\right]$. At each step, first, the search interval is halved using the midpoint $s$. Then, a feasibility test is performed to check whether the constraints C.1-C.4 of (P.1) can be met if the subset of MUs $\mathcal{X}^{[s]} \triangleq \left\{ \pi_{1}, \ldots, {{\pi_s}} \right\}$ is scheduled and the limits $\left[K_{\text{low}},K_{\text{up}}\right]$ is updated accordingly. The feasibility test is carried out by solving an instance of problem (\ref{P.4}) over the subset of MUs $ \mathcal{X}^{[s]}$. The feasibility status is determined by the value of the resulting auxiliary variables, i.e., the problem is considered to be feasible if the slack variables are smaller than a positive value $\eta$ close to zero.


We now discuss how to solve problem (\ref{P.4}). Problem (\ref{P.4}) is non-convex due to the non-convexity of the objective and of the constraints C.1 and C.2. Based on the limit $\|\mathbf{x}\|_0 =\lim_{\rm{p} \to 0} \|\mathbf{x}\|_{\rm p}^{\rm p} =\lim_{\rm p \to 0}\sum_{i_n\in \mathcal{X}} |x_{{i_n}}|^ {\rm p}$, the objective function of (\ref{P.4}) can be approximated by a higher-order norm to make the problem mathematically tractable.
In particular, in a manner similar to \cite{ran}, we adopt the smooth objective
\begin{equation} \label{smooth lp}
{\rm{f}} (\mathbf{x,y}) \triangleq \sum_{i_n\in \mathcal{X}} (x_{{i_n}}^2+\epsilon^2)^{{\rm p}/2}+\sum_{i_n\in \mathcal{X}} (y_{{i_n}}^2+\epsilon^2)^{{\rm p}/2},
\end{equation}
where $\epsilon > 0$ is a small fixed regularization parameter. Substituting (\ref{smooth lp}) as the objective in (\ref{P.4}), we now apply the SCA approach to obtain a local optimal solution of the resulting problem.

To this end, a convex upper bound satisfying conditions A1-A3 described in Sec. \ref{con_obj} for the smoothed ${\ell_p}$-norm objective function (\ref{smooth lp}) can be obtained from the result in \cite[Proposition 1]{ran} and is given by
\begin{equation} \label{smooth_lp_apprx}
{\tilde {\rm f}}\left( {{{\bf{Z}}}}; {{\bf{Z}}}\left( v \right) \right)  \triangleq {\sum_{i_n\in \mathcal{X}}  \omega'_{{i_n}} x_{{i_n}}^2}+{\sum_{i_n\in \mathcal{X}}  {\omega}''_{{i_n}} y_{{i_n}}^2}+ {\rm \bar f}\left( {{{\bf{Z}}}}; {{\bf{Z}}}\left( v \right) \right),
\end{equation} 
where ${\omega}'_{{i_n}}  \buildrel  \over =
         {\frac{\rm p}{2} \Big[  ({x}_{{i_n}}(v))^2 + \epsilon^2  } \Big]^{\frac{\rm p}{2}-1}$  and ${\omega}''_{{i_n}}  \buildrel  \over =
         {\frac{\rm p}{2} \Big[  ({y}_{{i_n}}(v))^2 + \epsilon^2  } \Big]^{\frac{\rm p}{2}-1}$; ${\bf{Z}}\left( v \right) \triangleq \big( {{{\bf{Q}}^{ul}}\left( v \right),{{\bf{Q}}^{dl}}\left( v \right)}, {{{\bf{f}}}\left( v \right),{{\bf{c}}^{ul}}\left( v \right)},{{{\bf{c}}^{dl}}\left( v \right)}, {{{\bf{x}}}\left( v \right)}, {{{\bf{y}}}\left( v \right)} \big)$; and The function ${\rm \bar f}\left( {{{\bf{Z}}}}; {{\bf{Z}}}\left( v \right) \right) \triangleq \frac{{{{{\gamma_{{q^{ul}}}}}}}}{2}\sum_{{i_n} \in {\cal I}}{\left\| {{{\bf{Q}}^{ul}} - {{\bf{Q}}^{ul}}\left( v \right)} \right\|^2}+\frac{{{{{\gamma_{{q^{dl}}}}}}}}{2}\sum_{{i_n} \in {\cal I}}{\left\| {{{\bf{Q}}^{dl}} - {{\bf{Q}}^{dl}}\left( v \right)} \right\|^2} + \frac{{{\gamma_f}}}{2}{\left\| {{\bf{f}} - {\bf{f}}\left( v \right)} \right\|^2} + \frac{{{\gamma_{{c^{ul}}}}}}{2}{\left\| {{{\bf{c}}^{ul}} - {{\bf{c}}^{ul}}\left( v \right)} \right\|^2} + \frac{{{\gamma_{{c^{dl}}}}}}{2}{\left\| {{{\bf{c}}^{dl}} - {{\bf{c}}^{dl}}\left( v \right)} \right\|^2} + \frac{{{\gamma_x}}}{2}{\left\| {{\bf{x}} - {\bf{x}}\left( v \right)} \right\|^2} + \frac{{{\gamma_y}}}{2}{\left\| {{\bf{y}} - {\bf{y}}\left( v \right)} \right\|^2} $ is added to realize the strong convexity of (\ref{smooth_lp_apprx}) with $\gamma_x, \gamma_y > 0$.

The convexification of constraint C.1 is done as in (\ref{g_t_apprx}).
 Lastly, to obtain an inner convexification for the energy constraint C.2 that satisfies the conditions B1-B6 in Sec. \ref{con_con}, we utilize the concave-convex structure of the rate function $r_{{i_n}}^{ul}\left( {{{\bf{Q}}^{ul}}} \right)$ as in (\ref{eq_6}), to rewrite constraint C.2 as
 \begin{equation}
 \text{tr}\left( {{\bf{Q}}_{{i_n}}^{ul}} \right)-\frac{E_{i_{n}}^M}{B^I_{{i_n}}} {r{{_{{i_n}}^{ul}}^ + }\left( {{{\bf{Q}}^{ul}}} \right)}-\frac{E_{i_{n}}^M}{B^I_{{i_n}}} {r{{_{{i_n}}^{ul}}^ - }\left( {{\bf{Q}}_{{-i_n}}^{ul}} \right)} \leq {y}_{{i_n}},
 \end{equation}
where ${r{{_{{i_n}}^{ul}}^ + }\left( {{{\bf{Q}}^{ul}}} \right)}$ and ${r{{_{{i_n}}^{ul}}^ - }\left( {{\bf{Q}}_{{-i_n}}^{ul}} \right)}$ are given in (18).
Using the linearization (\ref{g_t_apprx_2}), we then obtain the desired upper bound on C.2 as
 \begin{equation} \label{Ein_apprxx} 
   \text{tr}\left( {{\bf{Q}}_{{i_n}}^{ul}} \right)-\frac{E_{i_{n}}^M}{B^I_{{i_n}}} {\tilde{r}_{{i_n}}^{ul}\left( {{\bf{Q}}^{ul};{\bf{Q}}^{ul}\left( v \right)} \right)} \leq {y}_{{i_n}}.
\end{equation}
Given a feasible point 
 ${{\bf{Z}}\left( v \right)} $, we define the following strongly convex problem 
 \begin{align*}
& {\bf{\hat Z}}\left( {{\bf{Z}}\left( v \right)} \right) \triangleq \underset{\mathbf{Q}^{ul},\mathbf{Q}^{dl},\mathbf{f},\mathbf{c}^{ul},\mathbf{c}^{dl}, \mathbf{x},\mathbf{y}}{\text{argmin}}
   \begin{aligned}[t]
      &{\tilde {\rm f}}\left( {{{\bf{Z}}}}; {{\bf{Z}}}\left( v \right) \right)  \notag\\
   \end{aligned} \notag \\
   &  \;\;\;\;\;\;\;\; \;\;\;\;\;\;\;\; \;\;\;\;\;\;\;\; \;\;\;\;\;\;\;\text{s.t.} \notag \\
 & \begin{aligned}[t]
      & \mathbf{C.1}  \;\;\; {{\tilde g}_{{i_n}}}\left( {{{\bf{Q}}^{ul}},{{\bf{Q}}^{dl}};{{\bf{Z}}\left( v \right)}} \right) \notag+ \frac{{B_{{i_n}}^I}}{{c_{{i_n}}^{ul}{C^{ul}_n}}} + \frac{{B_{{i_n}}^O}}{{c_{{i_n}}^{dl}{C^{dl}_n}}}  \\
       & \;\;\;\;\;\;\;\; + \frac{{{V_{{i_n}}}}}{{{f_{{i_n}}}{F_c}}} - {T_{{i_n}}} \le x_{{i_n}}, \forall i_n \in \mathcal{X}, 
   \end{aligned}  \\ 
 & \mathbf{C.2}\;\; \; \text{tr}\left( {{\bf{Q}}_{{i_n}}^{ul}} \right)-\frac{E_{i_{n}}^M}{B^I_{{i_n}}} {\tilde{r}_{{i_n}}^{ul}\left( {{\bf{Q}}^{ul};{\bf{Q}}^{ul}\left( v \right)} \right)} \leq y_{i_{n}}, \forall i_n \in \mathcal{X},  \\
 {}&{{\bf{C}}{\bf{.3}}}\;\;\;x_{i_{n}} \geq 0, y_{i_{n}} \geq 0 , \forall i_n \in \mathcal{X}, \\
    & \mathbf{C.2}-\mathbf{C.4} \;\;\; \text {of (P.1)} , \forall i_n \in \mathcal{X}, \tag{P.5} \label{P.5}
      \end{align*}
where $
{\bf{\hat Z}}\left( {{\bf{Z}}} \right) \triangleq ({{{\bf{\hat Q}}}^{ul}},{{{\bf{\hat Q}}}^{dl}},{\bf{\hat f}},{{{\bf{\hat c}}}^{ul}}
,{{{\bf{\hat c}}}^{dl}},{{{\bf{ \hat x}}}},{{{\bf{ \hat y}}}})$ denote the unique solution of (\ref{P.5}). The SCA scheme for solving (\ref{P.5}) is described in Algorithm 3. As a technical note, we observe that here, since the approximant (\ref{smooth_lp_apprx}) of the objective function (\ref{smooth lp}) is an upper bound on (\ref{smooth lp}), convergence of Algorithm 3 is guaranteed also by setting $\gamma(v)=1$ \cite[Sec. III-A]{scutari2014distributed}.

\begin{algorithm}
\caption{User Scheduling}\label{alg:2}
\begin{algorithmic}[1]
\INPUT Parameters used by Algorithm 3.
\State Solve problem (\ref{P.4}) using Algorithm 3 for $\mathcal{X}=\mathcal{I}$ to obtain ${\bf{w}} \triangleq\dfrac{\bf{x}}{\sum_{{i_{n}\in\mathcal{I}}}{x_{i_{n}}}}+\dfrac{\bf{y}}{\sum_{{i_{n}\in\mathcal{I}}}{y_{i_{n}}}}$ and sort the MUs in ascending order of the value of $\mathbf{w}$ as  $ w_{\pi_{1}}\leq w_{\pi_{2}} \leq  \ldots \leq w_{{{\pi_{{KN_{c}}}}}} $, where $\pi$ is a permutation of $\mathcal{I}$.
    \State Set: $K_{\text{low}}=0$, $K_{\text{up}}=KN_c$.
    \State \textbf{Repeat} 
    \State Set $s \gets \floor{\frac{K_{\text{low}} + K_{\text{up}} }{2}}$.
    \State Perform the feasibility test by solving (\ref{P.4}) using Algorithm 3 for $\mathcal{X}=\mathcal{X}^{[s]} \triangleq \left\{ \pi_{1}, \ldots, {{\pi_s}} \right\}$: if it is feasible, set $K_{\text{low}}=s$; otherwise, set $K_{\text{up}}=s$.
        \State \textbf{Until} $K_{\text{up}}-K_{\text{low}}=1$.
         \State Set $s^{\ast}=K_{\text{low}}$ and $\mathcal{X}^{\ast} = \left\{ \pi_{1}, \ldots, {{\pi_{{s}^{\ast}}}} \right\}$.
             \OUTPUT Number of scheduled MUs $s^{\ast}$. 

\end{algorithmic}
\end{algorithm}
\begin{algorithm}
\caption{SCA Solution for (P.5)}\label{alg:3}
\begin{algorithmic}[1]
\INPUT Parameters from Table I; $ \rm p =0.5$; $v = 0$; ${\bf{Z}}\left( 0 \right) \in {\cal Z}$; ${\left\{ {\gamma \left( v \right)} \right\}_v} \in \left( {0,1} \right]$; ${{{\gamma_{{q^{ul}}}}}},{{\gamma_{{q^{dl}}}}},{\gamma_f},{\gamma_{{c^{ul}}}},{\gamma_{{c^{dl}}}},{\gamma_x},{\gamma_y},\epsilon > 0$; ${\omega'}_{{i_n}}(0)={\omega''}_{{i_n}}(0)= 1 $.
    \State If $\left| {{\rm \tilde{f}}\left( {{\bf{Z}};{{\bf{Z}}}\left( {v + 1} \right)} \right) - {\rm \tilde {f}}\left( {{\bf{Z}};{{\bf{Z}}}\left( v \right)} \right)}\right| \le \delta $, stop.
    
        \State Compute ${\bf{\hat Z}}\left( {{\bf{Z}}\left( v \right)} \right)$ from (P.5).
    \State Set ${\bf{Z}}\left( {v + 1} \right) = {\bf{Z}}\left( v \right) + \gamma \left( v \right)\left( {{\bf{\hat Z}}\left( {{\bf{Z}}\left( v \right)} \right) - {\bf{Z}}\left( v \right)} \right)$.
    \State Update
    \begin{align*}
&  {\omega'}_{{i_n}}(v+1)  \buildrel  \over =
   \begin{aligned}[t]
      &{\frac{p}{2} \Big[  ({x}_{{i_n}}(v+1))^2 + \epsilon^2  } \Big]^{\frac{p}{2}-1}, \\
   \end{aligned} \notag \\
   &  {\omega''}_{{i_n}}(v+1)  \buildrel  \over =
   \begin{aligned}[t]
      &{\frac{p}{2} \Big[  ({y}_{{i_n}}(v+1))^2 + \epsilon^2  } \Big]^{\frac{p}{2}-1}. \\
   \end{aligned} \notag 
   \end{align*}
    \State $v \leftarrow v + 1$, and return to step $1$. 
    \OUTPUT $\left({\mathbf{Q}^{ul},\mathbf{Q}^{dl},\mathbf{f},\mathbf{c}^{ul},\mathbf{c}^{dl}, \mathbf{x},\mathbf{y}}\right)$.
\end{algorithmic}
\end{algorithm}

%
%
\section{HYBRID EDGE and CLOUD COMPUTING} \label{sec:hybrid}
In the previous sections, we considered a scenario in which the MUs can offload applications to a cloud server. In this section, we extend the analysis to a more general set-up in which the ceNBs are directly connected to local computing servers, also known as cloudlets \cite{cloudlet}, which may run some of the MUs' applications. Specifically, each ceNB can either execute the computation task on the behalf of the MU or offload it to the cloud. Let ${F_n^{ceNB}}$ be the computation capability in CPU cycles per second of ceNB $n$, and let $f_{{i_n}}^{ceNB} \ge 0$ be the fraction of the ceNB's computing power assigned to user $i_n$, so that $\sum\limits_{i} {f_{{i_n}}^{ceNB}}  \le 1$. If implemented at the ceNB, the execution time of the task of MU $i_n$ is then given as
\begin{equation} \label{eq_3} 
\Delta _{{i_n}}^{exe|ceNB} = \frac{{{V_{{i_n}}}}}{{f_{{i_n}}^{ceNB}F_n^{ceNB}}}.
\end{equation}
%
In the same way, if the cloud processes the task of user $i_n$, the execution time $\Delta _{{i_n}}^{exe|cloud}$ is given by the right-hand side of (\ref{del_exe}).  The overall latency ${\Delta _{{i_n}}}$ experienced by each MU $i_n$ can be expressed as
\begin{equation} \label{eq_3} 
\begin{split}
{\Delta _{{i_n}}}  = \; & \Delta _{{i_n}}^{ul} + \left( {1 - {u_{{i_n}}}} \right)\Delta _{{i_n}}^{exe|ceNB} + {u_{{i_n}}}\Delta _{{i_n}}^{exe|cloud} \\
&+ {u_{{i_n}}}\Delta _{{i_n}}^{bh} + \Delta _{{i_n}}^{dl} ,
\end{split}
\end{equation}
where $\Delta _{{i_n}}^{ul}, \Delta _{{i_n}}^{bh}$ {and} $\Delta _{{i_n}}^{dl}$ have the same definition as in (\ref{del_ul}), (\ref{del_bh}) and (\ref{del_dl}), respectively; ${u_{{i_n}}}$ is a binary variable that indicates whether if the task of MU $i_n$ is processed on the ceNB (${u_{{i_n}}} = 0$) or on the cloud (${u_{{i_n}}} = 1$). 

To proceed, we relax the binary variable ${u_{{i_n}}}$ to be defined in the interval $\left[ {0,1} \right]$. This relaxation not only provides a lower bound on the minimum energy expenditure that can be obtained with a hard choice between cloudlet and cloud offloading, perhaps more importantly, it also captures a system in which the input data to the application of the MU $i_n$ can be split into two parts, of sizes  ${1 - {u_{{i_n}}}}$ and ${u_{{i_n}}}$, that can be processed separately at the ceNB and cloud, respectively. In the following, we will adopt this latter justification of the model.

As in Sec. \ref{sec:problem} and Sec. \ref{sec:sca}, we aim at minimizing the total energy consumed by the MUs to execute their tasks remotely under latency and power constraints. The problem is given by

   \begin{equation*}
\begin{array}{l}
\begin{array}{*{20}{l}}
{\mathop {{\mathop{\text{ min}}\nolimits} }\limits_{{{\bf{Q}}^{ul}},{{\bf{Q}}^{dl}},{\bf{u}},{\bf{f}^{ceNB}},{\bf{f}},{{\bf{c}}^{ul}},{{\bf{c}}^{dl}}} }&{E^{ul}\left( {{{\bf{Q}}^{ul}}} \right)}& { = \sum\limits_{{i_n} \in {\cal I}} {{E_{{i_n}}^{ul}}\left( {{\bf{Q}}_{{i_n}}^{ul},{\bf{Q}}_{-i_{n}}^{ul}} \right)} }  \\
{}&{}&{ = \sum\limits_{{i_n} \in {\cal I}} {{B_{{i_n}}^I}\frac{{\text{tr}\left( {{\bf{Q}}_{{i_n}}^{ul}} \right)}}{{r_{{i_n}}^{ul}\left( {{{\bf{Q}}^{ul}}} \right)}}} }
\end{array}\\
\begin{array}{*{20}{l}}
\;\;\;\;\;\;\;\;\;\;\;\;\;\;\;\; \text{s.t.}&{{\bf{C}}.\mathbf{1}}&{\frac{{B_{{i_n}}^I}}{{{W^{ul}}r_{{i_n}}^{ul}\left( {{{\bf{Q}}^{ul}}} \right)}} +\frac{{\left( {1 - {u_{{i_n}}}} \right){V_{{i_n}}}}}{{f_{{i_n}}^{ceNB}F_n^{ceNB}}}  }\\
{}&{}&{ +\frac{{u_{{i_n}}B_{{i_n}}^I}}{{c_{{i_n}}^{ul}{C^{ul}_n}}}+ \frac{{u_{{i_n}}{V_{{i_n}}}}}{{{f_{{i_n}}}{F_c}}}+\frac{{ u_{{i_n}} B_{{i_n}}^O}}{{c_{{i_n}}^{dl}{C^{dl}_n}}} }\\
 {}&{}& + \frac{{B_{{i_n}}^O}}{{{W^{dl}}r_{{i_n}}^{dl}\left( {{{\bf{Q}}^{dl}}} \right)}} \le {T_{{i_n}}}, \forall i_n \in \mathcal{I},\\

{}&{{\bf{C}}{\bf{.2}}}&{0 \le {u_{{i_n}}} \le 1}, \forall i_n \in \mathcal{I}, \;\;\;\;\;\;\;\;\;\;\;\;\;\;\;\;\;\;\;\; \\
{}&{{\bf{C}}{\bf{.3}}}& \resizebox{.55 \columnwidth}{!} 
{ ${f_{{i_n}}^{ceNB},f_{{i_n}}} \ge 0, \sum\limits_{{i_n} \in {\cal I}} {{f_{{i_n}}}}  \leq 1, \forall i_n \in \mathcal{I}  $},   \\
{}&{}& \sum\limits_{i=1}^K {f_{{i_n}}^{ceNB}}  \leq 1, \forall n=1,\ldots, N_c, \;\;\;\;\;\;\;\;\;\;\;\;\;\;\;\;\;\;\;\; \\
{}&{{\bf{C}}{\bf{.4}}}&c_{{i_n}}^{ul},c_{{i_n}}^{dl} \ge 0, \forall i_n \in \mathcal{I}, \\
 && \resizebox{.57 \columnwidth}{!} 
{$ \sum\limits_{i=1}^K {c_{{i_n}}^{ul}}  \leq 1, \sum\limits_{i=1}^K {c_{{i_n}}^{dl}}  \leq 1, \forall n=1,\ldots, N_c, $}\\
{}&{{\bf{C}}{\bf{.5}}}& {{\bf{Q}}_{{i_n}}^{ul} \in {\cal Q}_{{i_n}}^{ul}, {\forall i_n \in \mathcal{I}}}, \\
 {}&{}&{ \left(\mathbf{Q}_{i_{n}}^{dl}\right)_{i=1}^K \in {\cal Q}_n^{dl}, \forall n=1,\dots, N_c.}
\end{array}
\end{array}\notag \tag{P.6}\label{P.6}
\end{equation*} 
Note that, unlike (\ref{P.1}), here we have the additional optimization variables  ${\bf{u}} \triangleq {\left( {{u_{{i_n}}}} \right)_{{i_n} \in {\cal I}}}$ and ${\bf{f}}^{ceNB} \triangleq {\left( {f}^{ceNB}_{{i_n}} \right)_{{i_n} \in {\cal I}}}$.

It can be observed that \eqref{P.6} is non-convex due to the non-convexity of the objective function and constraint C.1. We tackle the problem by means of the SCA method convexifying the objective function as done in Sec. \ref{con_obj}. Furthermore, we need to calculate a convex upper bound for the C.1 constraint in (\ref{P.6}) that satisfies conditions B1-B6 in Sec. \ref{con_con}. Let us write the left-hand side of C.1 by
\begin{equation} \label{eq_g} 
\begin{split}
{g_{{i_n}}}&\left( {{{\bf{Q}}^{ul}},{{\bf{Q}}^{dl}},{\bf{u}},{{\bf{f}}^{ceNB}},{{\bf{f}}},{{\bf{c}}^{ul}},{{\bf{c}}^{dl}}} \right) \triangleq  \frac{{B_{{i_n}}^I}}{{{W^{ul}}r_{{i_n}}^{ul}\left( {{{\bf{Q}}^{ul}}} \right)}} \\
& + \frac{{B_{{i_n}}^O}}{{{W^{dl}}r_{{i_n}}^{dl}\left( {{{\bf{Q}}^{dl}}} \right)}} + \frac{{\left( {1 - {u_{{i_n}}}} \right){V_{{i_n}}}}}{{f_{{i_n}}^{ceNB}F_n^{ceNB}}}+ \frac{{{u_{{i_n}}}{V_{{i_n}}}}}{{f_{{i_n}}{F_c}}} \\
&  + \frac{{{u_{{i_n}}}B_{{i_n}}^I}}{{c_{{i_n}}^{ul}C_n^{ul}}} + \frac{{{u_{{i_n}}}B_{{i_n}}^O}}{{c_{{i_n}}^{dl}C_n^{dl}}}.
\end{split}
\end{equation}
To build the desired bound on ${g_{{i_n}}}$, we observe that the first two terms can be handled as in Sec. \ref{con_con}, while for the last four terms, we observe that they are all ratios of linear functions, we now observe that the relationship
\begin{equation} \label{eq_xy} 
\frac{x}{y} = \frac{1}{2}{\left( {x + \frac{1}{y}} \right)^2} - \frac{1}{2}\left( {{x^2} + \frac{1}{{{y^2}}}} \right),
\end{equation}
holds, where the right-hand side is the difference of two convex functions if $x \ge 0$ and $y > 0$. Therefore, a locally tight convex upper bound on the left-hand side of \eqref{eq_xy} can be obtained by linearizing the concave part as
\begin{equation} \label{eq_3} 
\begin{split}
\frac{x}{y} \le & \frac{1}{2}{\left( {x + \frac{1}{y}} \right)^2} - \frac{1}{2}\left( {{{\left( {{x^{v}}} \right)}^2} + \frac{1}{{{{\left( {{y^{v}}} \right)}^2}}}} \right) - {x^{v}}\left( {x - {x^{v}}} \right) + \\
& \frac{1}{{{{\left( {{y^{v}}} \right)}^3}}}\left( {y - {y^{v}}} \right)  ,
\end{split}
\end{equation}
where superscript $v$ identifies the point $\left(x^v, y^v \right)$ at which the upper bound is tight. Using (\ref{eq_3}) to the last four terms in (\ref{eq_g}), we define the desired approximants by substituting for $x$ and $y$ in (\ref{eq_3}) the numerator and denominator, respectively, of each of the last four terms in (\ref{eq_g}). This yields the SCA procedure in Algorithm 1 with the difference that we define  ${\bf{Z}}\left( v \right) \triangleq \left( {{{\bf{Q}}^{ul}}\left( v \right),{{\bf{Q}}^{dl}}\left( v \right),{\bf{u}}\left( v \right),{{\bf{f}}^{ceNB}}\left( v \right),{{\bf{f}}}\left( v \right),{{\bf{c}}^{ul}}\left( v \right),{{\bf{c}}^{dl}}\left( v \right)} \right)$ and ${\bf{Z}} \triangleq \left( {{{\bf{Q}}^{ul}},{{\bf{Q}}^{dl}},{\bf{u}},{{\bf{f}}^{ceNB}},{{\bf{f}}},{{\bf{c}}^{ul}},{{\bf{c}}^{dl}}} \right)$

\begin{align*}
& {\bf{\hat Z}}\left( {{\bf{Z}}\left( v \right)} \right) \triangleq \underset{\mathbf{Q}^{ul},\mathbf{Q}^{dl},, \mathbf{u}, \mathbf{f}^{ceNB},\mathbf{f},\mathbf{c}^{ul},\mathbf{c}^{dl}}{\text{argmin}}
   \begin{aligned}[t]
      &{\tilde E^{ul}\left( {{{\bf{Z}}}}; {{\bf{Z}}}\left( v \right) \right)   }\\
   \end{aligned} \notag \\
   & \;\;\;\;\;\;\;\;\;\;\;\;\;\;\;\;\;\;\;\;\; \;\;\;\;\;\;\;\;\;\;\;\;\;\text{s.t.} \notag \\
  & \mathbf{C.1}\;\; \; {{\tilde g}_{{i_n}}}\left( {\mathbf{Z}; {{\bf{Z}}}\left( v \right)   } \right)  \le {T_{{i_n}}}, \forall i_n \in \mathcal{I},\\
    & \mathbf{C.2}-\mathbf{C.5} \;\;\; \text {of (P.6)}, \tag{P.7} \label{P.7} 
     \end{align*}
where ${\bf{\hat Z}}\left( {{\bf{Z}}\left( v \right)} \right) \triangleq ({{{\bf{\hat Q}}}^{ul}},{{{\bf{\hat Q}}}^{dl}},{\bf{\hat u}},{{{\bf{\hat f}}}^{ceNB}},{{{\bf{\hat f}}}},{{{\bf{\hat c}}}^{ul}},{{{\bf{\hat c}}}^{dl}})$ denotes the unique solution of the strongly convex optimization problem (\ref{P.7}); the  objective function ${\tilde E^{ul}\left( {{{\bf{Z}}}}; {{\bf{Z}}}\left( v \right) \right)   } \triangleq \sum_{i_n \in \mathcal{I}} \Big( {\tilde E_{{i_n}}^{ul}}\left( {{{\bf{Q}}^{ul}}};{{{\bf{Q}}^{ul}}}(v) \right)+ {\bar E}_{{i_n}}\left( {{{\bf{Z}}_{{i_n}}}}; {{\bf{Z}}_{{i_n}}}\left( v \right) \right) \Big)$ where $  {\tilde E_{{i_n}}^{ul}}\left( {{{\bf{Q}}^{ul}}};{{{\bf{Q}}^{ul}}}(v) \right) $ is given as in  (\ref{Ein_apprx}) and we define  ${\bar E} \left( {{{\bf{Z}}_{{i_n}}}}; {{\bf{Z}}_{{i_n}}}\left( v \right) \right)\triangleq  \frac{{{\gamma_{{q^{ul}}}}}}{2}{\left\| {{{\bf{Q}}^{ul}_{{i_n}}} - {{\bf{Q}}^{ul}_{{i_n}}}\left( v \right)} \right\|^2} + \frac{{{\gamma_{{q^{dl}}}}}}{2}{\left\| {{{\bf{Q}}^{dl}_{{i_n}}} - {{\bf{Q}}^{dl}_{{i_n}}}\left( v \right)} \right\|^2} + \frac{{{\gamma_u}}}{2}{\left\| {{{u_{{i_n}}}} - {{u_{{i_n}}}}\left( v \right)} \right\|^2}$  $+$  $ \frac{{{\gamma_{{f^{ceNB}}}}}}{2}  \| {{{f}}^{ceNB}_{{i_n}}}$ - ${{{f}}^{ceNB}_{{i_n}}}\left( v \right)\|^2$ $+$ $ \frac{{{\gamma_{{f}}}}}{2}{\left\| {{{{f}}_{{i_n}}}  - {{{f}}_{{i_n}}}\left( v \right)} \right\|^2} + \frac{{{\gamma_{{c^{ul}}}}}}{2}{\left\| {{{{c}}_{{i_n}}^{ul}} - {{{c}}^{ul}_{{i_n}}}\left( v \right)} \right\|^2} + \frac{{{\gamma_{{c^{dl}}}}}}{2}{\left\| {{{{c}}^{dl}_{{i_n}}} - {{{c}}^{dl}_{{i_n}}}\left( v \right)} \right\|^2}$ with $\gamma_{{u}}, \gamma_{{f^{ceNB}}} > 0$; and ${{\tilde g}_{{i_n}}}\left( {\mathbf{Z}; {{\bf{Z}}}\left( v \right)   } \right)$ is the locally convex upper bound defined above.
%

%
\section{ENHANCED DOWNLINK VIA NETWORK MIMO} 
\label{netmimo}
%

So far, we  assumed that each ceNB $n$ serves the users $\left\{i_n:  i=1,\ldots,K\right\}$ in its cell, hence interfering with other ceNBs. In this section, instead, we consider enhanced downlink transmission based on network MIMO. Specifically, we assume that each MU can receive the result of the cloud execution in the downlink not only from ceNB in the same cell, but also from other ceNBs that cooperatively transmit to the MU. Cooperation among ceNBs is enabled by the transmission on the backhaul links of the outcome of the cloud execution for a given MU to multiple ceNBs. We specifically consider here the extreme case of full ceNB cooperation, so that all the ceNBs transmit cooperatively to each MU. The extension to a more general model with clustered cooperation is straightforward and will not be pursued further.

To express the achievable downlink rate with network MIMO, it is convenient to define user-centric transmit covariance matrices ${\mathbf{Q}}^{dl}_j \in \mathbb{C}^{ N_{{T}}\times N_{{T}} }$ for every MU $j \in \cal{I}$, where we have defined $N_T \triangleq \sum_{n=1}^{N_c} N_{{T}_{n}} $, such that the $n$-th $N_{{T}_{n}}\times N_{{T}_{n}} $ block on the main diagonal, denoted by $[{\bf{Q}}_j^{dl}]_n$, represents the contribution of ceNB $n$ to the transmission to MU $j$. Note that the out-of-diagonal blocks in ${\bf{Q}}_j^{dl}$ describe the correlation among signals sent by different ceNBs, which designates cooperative transmission at the ceNBs. The set of downlink covariance matrices, ${\mathbf{Q}}^{dl}= \left\{{{\mathbf{Q}}_{{j}}^{dl}}\right\}_{ j \in \cal{I}} $  is
\begin{equation} 
\begin{split}
\mathcal{Q}_{{}}^{dl}  \triangleq \big\{  {{\mathbf{Q}}_j^{dl} \in \mathbb{C}^{N_{{T}}\times N_{{T}} }, \;  j \in \cal{I} } :   \displaystyle\sum_{j \in \cal{I}} \text{tr}  \big( [{\bf{Q}}_j^{dl}]_n \big) & \leq P_n^{dl}, \forall n  \big\},
\end{split}
\end{equation}
so that the per-ceNB power constraints are satisfied. Also, it is convenient to define the channel matrix from all the ceNBs to MU $j$ as ${\tilde{\bf{G}}_{{j}}} \in {\mathbb{C}^{N_{{R_j}} \times {N_{{T}}}}}$, where  ${\tilde{\bf{G}}_{{j}}} \triangleq [{\bf{G}}_{{j}1},  {\bf{G}}_{{j}2},\ldots, {\bf{G}}_{{j}N_c}]$ and we have set ${{\bf{G}}_{{j},n}}={{\bf{G}}_{{j}}}$. With these definitions, the achievable downlink rate is given by
\begin{equation}
r_{i_{n}}^
{dl}(\mathbf{Q}^
{dl})=\log_{2}{\det}\Big(\mathbf{I}+ \tilde{\mathbf{G}}_{i_{n}}^H  \tilde{\mathbf{R}}_{i_{n}}^{dl}(\mathbf{Q}_{-i_{n}}^
{dl})^{-1}\tilde{\mathbf{G}}_{i_{n}}{\mathbf{Q}}_{i_{n}}^{dl}\Big),
\end{equation}
with
\begin{equation}
\tilde{\mathbf{R}}_{i_{n}}^{dl}(\mathbf{Q}_{-i_{n}}^
{dl})\triangleq\sigma_w^2\mathbf{I}+ \sum_{i_m \in {\cal I}\backslash \left\{ {{i_n}} \right\}}\tilde{\mathbf{G}}_{i_{n}}\mathbf{Q}_{i_{m}}^{dl}\tilde{\mathbf{G}}_{i_{n}}^H,
\end{equation}
and $\mathbf{Q}_{-i_{n}}^{dl}\triangleq(\mathbf{Q}_{j_{m}}^{dl})_{j_m\ne i_n}$.

In order to enable network MIMO, we need to ensure that downlink transmissions from all ceNBs take place at the same time. To this end, we impose that uplink transmission, computing and backhaul transmissions for all MUs are constrained to be completed by a given time $T_1$. At time $T_1$, then, the downlink transmission is initiated and takes a given time $T_2$. Accordingly, we formulate the  optimization problem following the weighted sum approach discussed in Remark 2 as
\begin{equation*}
\begin{array}{l}
\begin{array}{*{20}{l}}
{\mathop {{\mathop{\rm min}\nolimits} }\limits_{{{\bf{Q}}^{ul}},{{\bf{Q}}^{dl}},{\bf{f}},{{\bf{c}}^{ul}},{{\bf{c}}^{dl}},T_1,T_2} }&&{ \sum\limits_{{i_n} \in {\cal I}} {{B_{{i_n}}^I}\frac{{\text{tr}\left( {{\bf{Q}}_{{i_n}}^{ul}} \right)}}{{r_{{i_n}}^{ul}\left( {{{\bf{Q}}^{ul}}} \right)}}}+\lambda (T_1 + T_2) }\\
\end{array}\\
\begin{array}{*{20}{l}}
\;\;\;\;\;\;\;\;\;\;\;\;\; \text{s.t.}&{{\bf{C}}.\mathbf{1}}&{\frac{{B_{{i_n}}^I}}{{{W^{ul}}r_{{i_n}}^{ul}\left( {{{\bf{Q}}^{ul}}} \right)}} + \frac{{B_{{i_n}}^I}}{{c_{{i_n}}^{ul}{C^{ul}_n}}} + \frac{{{V_{{i_n}}}}}{{{f_{{i_n}}}{F_c}}}}\\
{}&{}&{ + {\frac{{B_{{i_n}}^O}}{{C_m^{dl}c_{{i_n},m}^{dl}}}}  \le {T_1}}, \forall i_n \in \mathcal{I},  \\
{}&{\bf{C.2}}&{ \frac{{B_{{i_n}}^O}}{{{W^{dl}}r_{{i_n}}^{dl}\left( {\bf{Q}}^{dl} \right)}}  \le {T_2}}, \forall i_n \in \mathcal{I},\\
{}&{{\bf{C}}{\bf{.3}}}&{{f_{{i_n}}} \ge 0,\sum\limits_{{i_n} \in {\cal I}} {{f_{{i_n}}}}  \leq 1}, \forall i_n \in \mathcal{I}, \;\;\;\;\;\;\;\;\;\;\;\;\;\;\;\;\;\;\;\; \\
{}&{\bf{C.4}}&c_{{i_n}}^{ul},c_{{i_n},m}^{dl} \ge 0,\sum\limits_{i=1}^K {c_{{i_n}}^{ul}}  \le 1, \\ 
{}&{}&\sum\limits_{{j} \in {\cal I}} {c_{{j},m}^{dl}}  \le 1, \forall i_n \in \mathcal{I},\\
{}&{{\bf{C}}{\bf{.5}}}&{{\bf{Q}}_{{i_n}}^{ul} \in {\cal Q}_{{i_n}}^{ul},{\bf{Q}}^{dl} \in {\cal Q}^{dl}}, \forall i_n \in \mathcal{I},
\end{array}
\end{array}\notag \tag{P.8}\label{P.8}
\end{equation*} 
where ${{\bf{c}}^{dl}} \triangleq {\left( {c_{{i_n},m}^{dl}} \right)_{{i_n} \in {\cal I},m}}$, with ${c_{{i_n},m}^{dl}}$ being the fraction of the backhaul capacity to ceNB $m$ allocated to transmit the output bits intended for MU $i_n$; and $\lambda >0$ is a parameter defining the relative weight of energy and latency. 

Problem (\ref{P.8}) is non-convex due to the non-convexity of the objective function and constraints C.1 and C.2. We tackle this problem using SCA in  a way similar to that used for (P.1), i.e., we obtain a convex approximants for the objective functions using (\ref{Ein_apprx}) and for the latency constraints using (\ref{g_t_apprx}{}). The problem is then solved using the SCA procedure described in Algorithm 1.

\section{Numerical Results}

In this section, we present numerical results validating the model and algorithms  presented in the previous sections. Throughout, we consider a network composed of three cells with five users in each cell, i.e., $N_c=3$ and $K=5$. All transceivers are equipped with $N_{T_{i_n}}=N_{R_{n}}=2$ antennas. The channel matrices are generated with independent and identically distributed complex Gaussian entries having zero mean and variance equal to the path loss, which is assumed to be identical for uplink and downlink. The path loss is given by 170 dBm between an MU and the ceNB in the same small cell and 180 dBm between an MU and the ceNB in the other small cell. The values 170 dBm and 180 dBm can be justified by considering the Walfish-Ikegami model \cite {loss} with distances of $500$ meters between MU and ceNB in the same cell and $700$ meters for MU and ceNB in different cells. The other  parameters of the Walfish-Ikegami model are selected so as to simulate a small-cell environment in a typical urban setting as listed in Table II. Targeting a small-cell scenarios, we will explore values of the backhaul capacities as low as few Mbits/s \cite{g5,bh}.  Finally, note that setting $N_0=-170$ dBm/Hz \cite{loss}, yields an average signal-to-noise ratio (SNR) on the direct link of $40$ dBm per receive antenna and $30$ dBm on the interference link for a signal transmitted at $0.01$ Joule per symbol from a single antenna in the uplink and downlink. Furthermore, unless stated otherwise, we select each number of input bits $B_{i_{n}}^I$ and output bits $B_{i_{n}}^O$  uniformly at random in the interval $[0.1-1]$ Mbits and  we set the number of CPU cycles as $V_{i_{n}}=2640\times B_{i_{n}}^I$ CPU cycles. These choices reflect computational-intensive applications, as demonstrated by the measurments in \cite{Miettinen2010EE,emill2011app}. The cloud capacity is $F_c=10^{11}$ CPU cycles/s, which is, e.g., obtained by a four-core server with Intel Xeon processor with 3.3 GHz that is commercially used by Amazon elastic compute cloud (EC2) \cite{ec2,kumar2010cloud}. Other system parameters are set to $W^{ul} = W^{dl} = 10$ MHz, $C_n^{ul} = C_n^{dl} = 100$ Mbits/s, $d_{i_{n}}=10^{-5}$ J/symbol \cite{d2009},  and $T_{i_{n}}=0.1$ seconds. Throughout, averages are intended with respect to the channel realizations.
\begin{table}[ht!]
\begin{minipage}{\columnwidth}
\centering
\begin{tabular}{|p{1cm}|p{0.5cm}|p{5cm}|}
\hline

parameter      & value & description \\  \hline

$f$ & $1800$       & frequency (MHz) \\ \hline

$\theta$ & $45 ^{\circ}$       & road orientation angle (degree) \\ \hline

$h_\text{Tx}$  &  $3$    & height of transmitter (meters) \\ \hline
$h_\text{Rx}$  &  $1$    & height of receiver (meters) \\ \hline
$h_\text{Roof}$&  $5.5$  & mean value of buildings height (meters) \\ \hline
$s_\text{Roof}$&  $8$    & mean value of buildings separation (meters) \\ \hline
str\_wid&  $5$    & mean value of street width (meters) \\ \hline
$k_a$   &  $54$    & path loss penalty when ceNB below rooftop \\ \hline
$k_d$   &  $5$    & correlation adjustment  constant \\
\hline
\end{tabular}
\caption{Parameters in the Walfish-Ikegami path loss model \cite {loss}.}
\label{t-ex1}
\end{minipage} 
\end{table}
\begin{figure}[!t]
        \centering
        \includegraphics[width=\columnwidth]{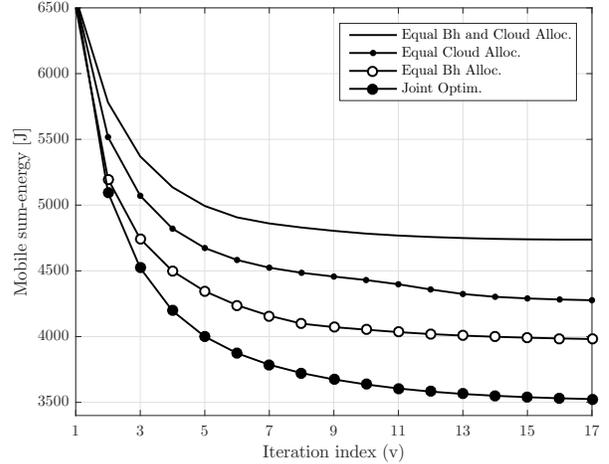}
        \caption{{\small{Minimum average mobile energy consumption versus iteration index ($T_{i_n}=0.12$ seconds, $N_c=3, K=5$, $W^{ul}=W^{dl}=100$ MHz,  $B_{{i}_n}^I$ and $B^O_{{i}_n} \sim \mathcal{U}\{0.1, 1\}$ Mbits, $V_{i_{n}}=2640\times B_{i_{n}}^I$ CPU cycles, $C_{n}^{ul}= C_{n}^{dl}=100$ Mbits/s, and $F_c=10^{11}$ CPU cycles/s).}}}
        \label{fig1}
\end{figure}
%

We start by illustrating the typical convergence properties of the SCA algorithm by plotting in Fig. 3 the average minimal mobile energy consumption ${E }\left( {{{\bf{Q}}^{ul}},{{\bf{Q}}^{dl}}} \right)$ versus the iteration index $v$. Besides the joint optimization across uplink, downlink, backhaul and computing resources, studied in Sec. \ref{sec:problem}, we consider the following more conventional solutions: \emph{(i) Equal backhaul and cloud allocation:} the computing and backhaul resources are equally allocated to all MUs, that is, $f_{i_n}=1/(N_cK)$ and $c^{ul}_{i_n}=c^{dl}_{i_n}=1/K$ for all $i_n \in \cal{I}$, while the covariance transmit and receive matrices at the physical layer are optimized using SCA \cite{sar02}; \emph{(ii) Equal cloud allocation:} computing resources at the cloud are equally allocated among the MUs, while the rest of the parameters are jointly optimized using SCA; \emph{(iii) Equal backhaul allocation:} the backhaul resources are equally allocated among the MUs, while the rest of the parameters are jointly optimized using SCA. We observe the fast convergence of SCA, which requires around 17 iterations to obtain results close to convergence (at iteration 17 the termination criterion $\left| {E\left( {{{\bf{Q}}^{ul}}\left( {v + 1} \right)}, {{{\bf{Q}}^{dl}}\left( {v + 1} \right)} \right) - E\left( {{{\bf{Q}}^{ul}}\left( v \right)}, {{{\bf{Q}}^{dl}}\left( v \right)} \right)} \right| \le \delta $ is satisfied with $\delta=10^{-3}$ for $\alpha = 10^{-5}$). Furthermore, it can be seen that the proposed joint optimization method shows a considerable gain compared to the equal allocation of computational and backhaul resources.

The gains that can be accrued by means of joint optimization are further investigated in Fig. 4 (obtained in the same setting of Fig. 3), which depicts the minimum average mobile energy as a function of the latency constraints $T_{i_{n}}$, which are assumed to be the same for all MUs. The energy saving due to joint optimization can be seen to be particularly pronounced in the regime in which the latency constraint is more stringent. For instance, at $T_{i_{n}}=0.12$ seconds, which is the smallest latency for which all schemes are feasible, joint optimization saves 66\% in terms of sum-energy as compared to equal allocation of backhaul and cloud resources, 48\% as compared to equal cloud allocation, and 32\% as compared to equal backhaul allocation. 
\begin{figure}[!t]
        \centering
        \includegraphics[width=\columnwidth]{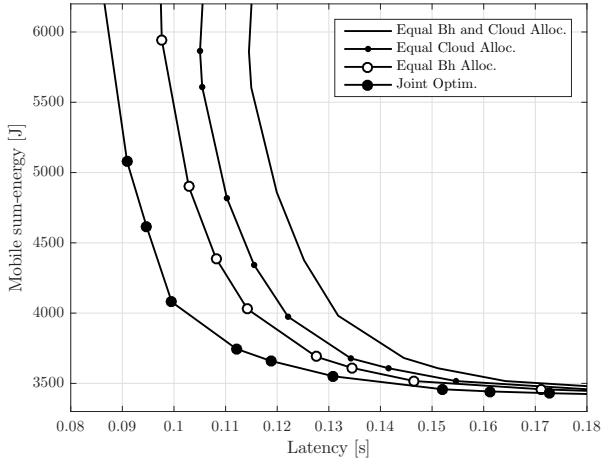}
        \caption{{\small{Minimum average mobile energy consumption versus the latency constraint $T_{i_{n}}$, assumed to be the same for all MUs ($N_c=3, K=5$, $W^{ul}=W^{dl}=100$ MHz,  $B_{{i}_n}^I$ and $B^O_{{i}_n} \sim \mathcal{U}\{0.1, 1\}$ Mbits, $V_{i_{n}}=2640\times B_{i_{n}}^I$ CPU cycles, $C_{n}^{ul}= C_{n}^{dl}=100$ Mbits/s, and $F_c=10^{11}$ CPU cycles/s). }}}
        \label{fig2}
\end{figure}
%

To account for the case in which the network may operate under asymmetrical bandwidth allocation for uplink and downlink, we compare the mobile sum-energy consumption of the schemes considered above in the same set-up of Fig. 3 and Fig. 4, with the caveat that we set $W^{dl}=10$ MHz and we vary the uplink/downlink bandwidth ratio $W^{ul}/W^{dl}$, in Fig. 5. We observe that joint optimization is especially advantageous in term of mobile energy consumption when the uplink bandwidth is more constrained than the downlink bandwidth. This is because, in this regime, it is particularly useful to allocate more computing and backhaul resources to users with worse channel condition in order to meet the latency requirements with minimal energy expenditure. For example, when $W^{ul}$ is five times smaller than the downlink bandwidth, the joint optimization scheme is $50\%$ more energy efficient than the fixed allocation of backhaul and cloud resources.

\begin{figure}[h] \centering
  \includegraphics[clip,width=\columnwidth]{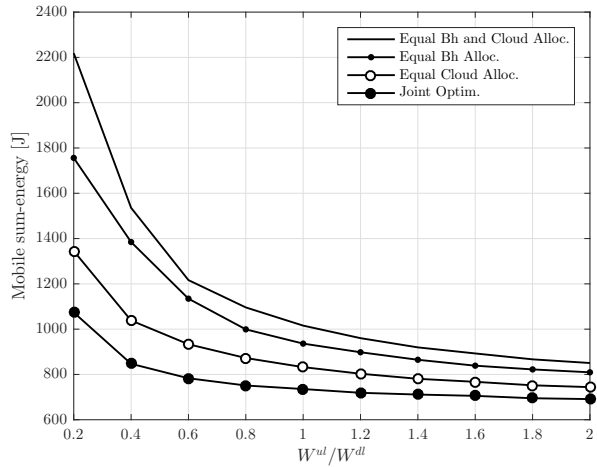}%
\centering
\caption{\small{ Minimum average mobile sum-energy consumption versus bandwidth allocation ratio $W^{ul}/W^{dl}$ ($W^{dl}=10$ MHz, $N_c=2, K=2, T_{i_{n}}=0.09$ s, $B_{{i}_n}^I$ and $B^O_{{i}_n} \sim \mathcal{U}\{0.1, 1\}$ Mbits, $V_{i_{n}}=2640\times B_{i_{n}}^I$ CPU cycles, $C_{n}^{ul}= C_{n}^{dl}=100$ Mbits/s, and $F_c=10^{11}$ CPU cycles/s).}}
\end{figure}

While the previous results assume that all MUs perform computation offloading, we next turn to the study of user selection that is presented in Sec. IV. To account for mobile energy consumption, we set the effective switched capacitance of the mobile as $\kappa=10^{-26}$ so that the mobile energy consumption is consistent with the measurements made in \cite{Miettinen2010EE} for a Nokia N900 mobile device operating at frequency 500 MHz.  To this end, we plot the average number of selected MUs as a function of the latency constraint $T_{i_{n}}$, assumed to be the same for all MUs, for different values of backhaul in Fig. 6. The figure shows the results obtained by means of the efficient algorithm proposed in Sec. IV with $\eta=10^{-5}$, as well as by an \textit{exhaustive search} strategy in which, for every subset of users, a joint optimization problem is solved as per Sec. II using SCA (Algorithm 1) and then the subset of users which yields the minimal energy consumption is selected. The figure demonstrates how a less restrictive latency constraint and/or a larger backhaul increases the number of users that should be allowed to offload. For instance, for $T_{{i_n}} \leq 0.07$ seconds, there is no MU on average offloads its applications  since offloading is more energy demanding. Instead, for $T_{i_n} > 0.2$ seconds, as long as the backhaul capacity is larger than $500$ Mbits/s, all MUs tend to offload.  It is also observed that the proposed scheme selects a number of users close to that chosen by exhaustive search.
\begin{figure}[!t]
        \centering
        \includegraphics[width=\columnwidth]{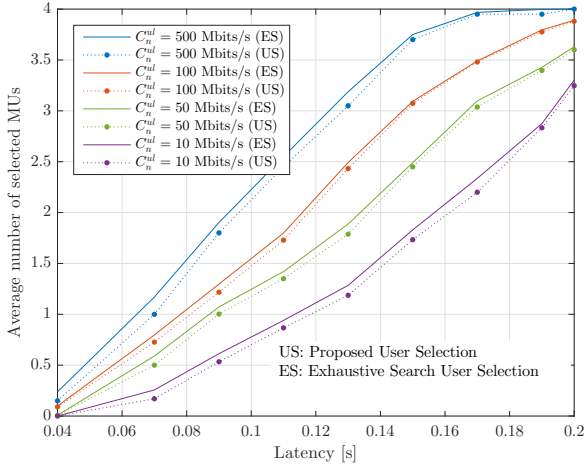}
        \caption{{\small{ Average number of selected MUs versus latency constraint $T_{i_{n}}$ for both the proposed efficient scheme in Sec. IV and exhaustive search ($N_c=2, K=2$, $B_{i_{n}}^I=B_{i_{n}}^O=1 $ Mbits, $V_{i_{n}}=10^9$ CPU cycles, $F_c=10^{11}$ CPU cycles/s, and $C_{n}^{dl}= C_{n}^{ul}$).}}}
        \label{fig5}
\end{figure}

The corresponding overall energy consumption for the example of Fig. 6 with backhaul capacity $C_{n}^{ul}= C_{n}^{dl}=100$ Mbits/s is shown in Fig. 7. Note that the overall energy consumption is the sum of the mobile computing energy for the MUs that perform local computing, namely (\ref{local}), and of the energy used for transmission, namely (\ref{off_ene}), for the MUs that offload. For reference, we also show the energy required when all MUs perform their tasks locally. The proposed user selection scheme is seen to achieve a near-optimal energy performance compared to the exhaustive search method. As an example, when the latency constraint $T_{i_{n}} =0.17 $ seconds, around 83\% energy saving can be obtained with the proposed user offloading selection scheme as compared to \textit{local computing}.
\begin{figure}[!t]
        \centering
        \includegraphics[width=\columnwidth]{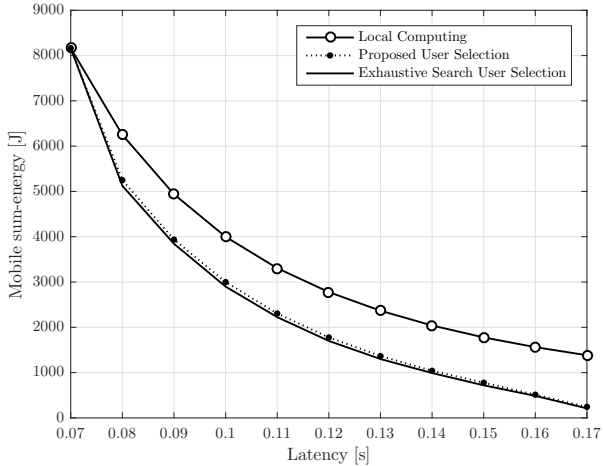}
        \caption{{\small{Minimum average mobile energy consumption versus the latency constraint $T_{i_{n}}$ ($N_c=2, K=2$, $B_{i_{n}}^I=B_{i_{n}}^O=1 $ Mbits, $V_{i_{n}}=10^9$ CPU cycles, $C_{n}^{ul}= C_{n}^{dl}=100$ Mbits/s, and $F_c=10^{11}$ CPU cycles/s).}}}
        \label{fig5}
\end{figure}

In the previous results, central cloud processing was assumed. We now investigate the optimal allocation of computing tasks between the edge and the cloud in the set-up studied in Sec. V. To this end, in Fig. 8, we plot the fractions ${\left( {{{{{\hat u}}}_{{i_n}}}} \right)_{{i_n} \in {\cal I}}}$ of the application of each MU $i_n$ to be performed at the cloud as a function of the backhaul capacity when the computing capability of the cloudlet is given by  $F_n^{ceNB} = 0.1 {F_c}$. To highlight the impact of asymmetries in the offloading requirements, we assume here that in the first cell the two users have $B_{{1_1}}^I = B_{{1_1}}^O = 1$ Mbits, $   {V_{{1_1}}} = {10^9}$ CPU cycles, and $B_{{2_1}}^I = B_{{2_1}}^O = 0.7$ Mbits, ${V_{{2_1}}} = 0.7{V_{{1_1}}}$ CPU cycles, while the users in the second cell have smaller offloading requirements for data transfer and computing, normally $B_{{1_2}}^I = B_{{1_2}}^O = 0.5$ Mbits, ${V_{{1_2}}} = 0.5{V_{{1_1}}}$ CPU cycles, and $B_{{2_2}}^I = B_{{2_2}}^O = 0.1$ Mbits, ${V_{{2_2}}} = 0.1{V_{{1_1}}}$ CPU cycles. It is observed that, when the backhaul capacity is limited, tasks tends to be performed at the edge, especially for users with more stringent data transfer and computing requirements, for example, at $C_{n}^{ul}= C_{n}^{dl}=3$ Mbits/s, the two users in the first cell execute $50\%$ and $80\%$ percent of their offloaded tasks at the cloud, while the users in the second cell have a complete tasks execution at the cloud due to the moderate number of input/output bits. As the backhaul capacity is increased, more tasks are executed at the cloud, benefiting from the improved connection between ceNBs and the cloud.
\begin{figure}[!t]
        \centering
        \includegraphics[width=\columnwidth]{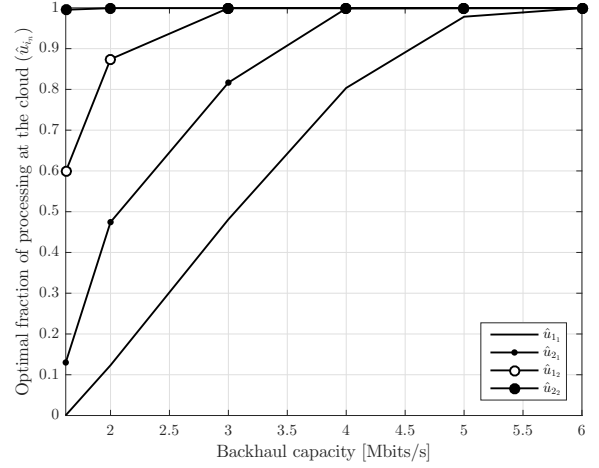}
        \caption{{\small{Offloading cloud vs. cloudlet indicator versus the backhaul capacity ($N_c=2, K=2$, $T_{i_{n}}=0.9$ seconds, $B_{{1_1}}^I = B_{{1_1}}^O = 1$ Mbits, $B_{{2_1}}^I = B_{{2_1}}^O = 0.7$ Mbits, $B_{{1_2}}^I = B_{{1_2}}^O = 0.5$ Mbits, $B_{{2_2}}^I = B_{{2_2}}^O = 0.1$ Mbits; ${V_{{1_1}}} = {10^9}$ CPU cycles, ${V_{{2_1}}} = 0.7{V_{{1_1}}}$, ${V_{{1_2}}} = 0.5{V_{{1_1}}}$ and ${V_{{2_2}}} = 0.1{V_{{1_1}}}$, $C_{n}^{ul}= C_{n}^{dl}=100$ Mbits/s, $F_n^{ceNB} = 10^{10}$ CPU cycles/s, and  $F_c=10^{11}$ CPU cycles/s).}}}
        \label{fig5}
\end{figure}

\begin{figure}[h] \centering
  \includegraphics[clip,width=\columnwidth]{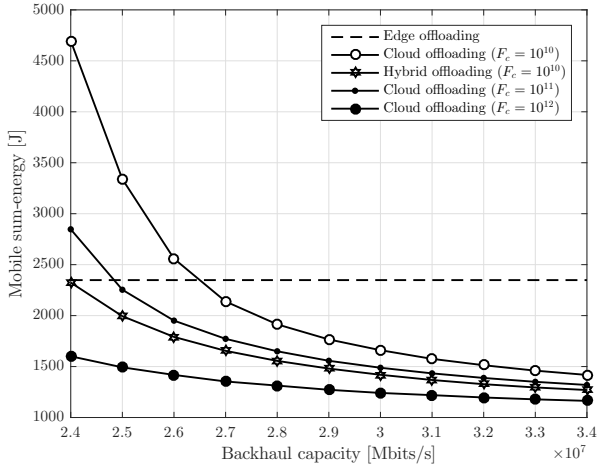}%
\centering
\caption{\small{ {Minimum average mobile sum-energy consumption for cloud and edge offloading, along with the proposed hybrid edge-cloud offloading scheme, versus backhaul capacity $C_n^{ul}=C_n^{dl}$ ($N_c=2, K=2, T_{i_{n}}=0.1$ s, $W^{ul}=W^{dl}=10$ MHz, $B_{{i}_n}^I$ and $B^O_{{i}_n} \sim \mathcal{U}\{0.1, 1\}$ Mbits, $V_{i_{n}}=2640\times B_{i_{n}}^I$ CPU cycles, and $F_n^{\text{ceNB}}=10^{10}$ CPU cycles/s).}}}
\end{figure}

To obtain further insights, Fig. 9. shows the mobile sum-energy consumption versus the backhaul capacity for cloud mobile computing, whereby all users offload to the cloud, and edge mobile computing, whereby all users offload to the local cloudlet, for $N_c=2, K=2$, and $F_n^{\text{ceNB}}=10^{10}$ CPU cycles/s. We also include the performance of the solution introduced in Sec. V, which allows for hybrid edge-cloud offloading. For cloud offloading, we consider different values for computational capacity $F_c$ of the cloud. It is seen that when the backhaul capacity constraint is stringent, edge offloading is more energy efficient as compared to cloud offloading, e.g., by about 44\% for cloud computational capacity $F_c=10^{10}$ CPU cycles/s and backhaul capacity $C_n^{ul}=C_n^{dl} = 2.4 $ Mbits/s. Furthermore, larger values of cloud capacity $F_c$ can compensate for limited backhaul resources, making cloud computing preferable to edge computing. It is also observed that the hybrid edge-cloud offloading scheme can significantly outperform edge computing when backhaul capacity is limited.

We finally turn to the performance of the downlink network MIMO scheme presented in Sec. VI. In Fig. 10, we plot the average mobile sum-energy with and without cooperative transmission in the downlink. The performance of cooperative scheme is obtained from the solution of (P.8), while the performance of non-cooperative scheme is obtained from the solution of (P.1) under the indicated values of backhaul capacity. The users offloading requirements are identical to that used in Fig. 6. The key observation here is that the performance of the downlink cooperative transmission is strongly limited by the backhaul capacity. For instance, we can see that, with backhaul $C_{n}^{ul}= C_{n}^{dl}=10$ Mbits/s, the cooperative scheme is about $85\%$ more energy-consuming as compared to the non-cooperative scheme at latency $0.5$ seconds. The energy performance gap between the two schemes is diminished as the backhaul increases as observed with $C_{n}^{ul}= C_{n}^{dl}=100$ Mbits/s. With this value of backhaul capacity, the cooperative scheme is less energy efficient by $45\%$ around $T_{i_{n}}=0.3$ seconds. However, with backhaul $C_{n}^{ul}= C_{n}^{dl}=10$ Gbits/s, as for a standard fiber optic channel, the cooperative scheme starts to have noticeable energy gains. At latency $0.08$ seconds, for example, the cooperative scheme attains $57\%$ energy saving as compared to the non-cooperative scheme.

\begin{figure}[!t]
        \centering
        \includegraphics[width=\columnwidth]{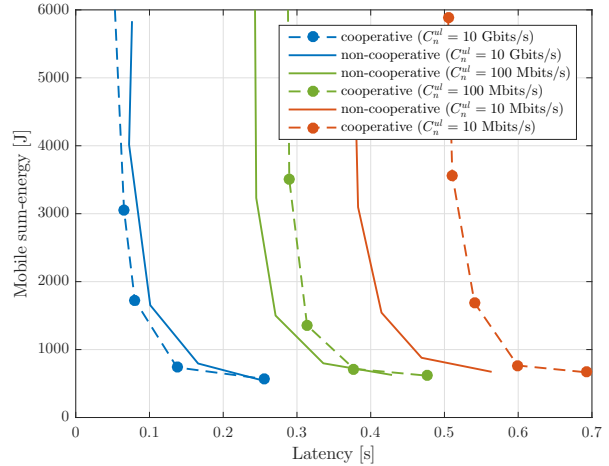}
        \caption{{\small{Minimum average mobile energy consumption versus the latency constraint $T_{i_{n}}$, assumed to be the same for all MUs ($N_c=2, K=2$, $B_{1_1}^I=B_{1_2}^I=1$ Mbits, $B_{2_1}^I=B_{2_2}^I=0.1$ Mbits, $B_{1_1}^O=B_{1_2}^O=1 $ Mbits, $B_{2_1}^O=B_{2_2}^O=0.1$ Mbits, $V_{1_1}= V_{1_2}=10^9$ CPU cycles, $V_{2_1}= V_{2_2}=10^8$ CPU cycles, $F_c=10^{11}$ CPU cycles/s, and $C_{n}^{dl}= C_{n}^{ul}$).}}}
        \label{fig6}
\end{figure}

\section{CONCLUDING REMARKS}
In this paper, we investigated the design of cloud mobile computing systems over  MIMO cellular networks as a joint optimization problem over radio, computational resources and backhaul resources in both uplink and downlink directions. An iterative algorithm based on successive convex approximations was presented for solving the resulting non-convex problem under latency and power constraints. Numerical results show that the proposed joint optimization yields significant energy saving compared to the conventional solutions based on separate allocations of computing and/or backhaul resources. This saving is more pronounced in low latency regimes, where, in our results, it leads to energy saving as high as $66\%$. The mentioned baseline problem was further generalized in several directions. First, we tackled user selection with the aim of ensuring an energy savings for all users that perform offloading. Then, a hybrid architecture  that leverages both edge and cloud  computing was studied by addressing the optimal  allocation between cloud and edge.  It was seen that the optimal allocation is significantly affected by the backhaul capacity. Finally, joint downlink transmission based on network MIMO was considered, demonstrating the critical importance of the backhaul for the viability of this technique. Among the problems left open by this study, we mention here the consideration of higher-layer aspects such as queuing in the definition of the latency.

\bibliographystyle{ieeetran}
\bibliography{refs}

\end {document}